\documentclass[aps,pra,floatfix,twocolumn,superscriptaddress]{revtex4-2}
\usepackage{graphicx,
            caption,
            subcaption,
            amsmath,
            braket,
            amssymb,
            amsfonts,
            mathtools,
            ragged2e,
            xspace,
            nicefrac,
            lipsum,
            nameref,
            textcomp,
            urwchancal,
            float,
            upgreek,
            isotope,
            hyperref,
            cleveref,
            siunitx,
            placeins, % for \FloatBarrier
            scrextend, % for KOMA script functionalities
            url}

\usepackage{gensymb}

\usepackage{scalerel}
\usepackage{tikz}
\usetikzlibrary{svg.path}

\DeclareCaptionLabelSeparator{dot}{. }
\makeatletter
\def\justified{
	\let\\\@normalcr
	\@rightskip\z@skip \rightskip\@rightskip
	\leftskip\z@skip
	\parindent 0em\relax
	\setlength{\parfillskip}{0pt plus 1fil}}
\DeclareCaptionJustification{justified}{\justified}
\hypersetup{colorlinks=true, linkcolor=blue,citecolor=blue,urlcolor=blue}
\def\unit #1 #2 {\SI{#1}{#2}\xspace}
\def\range #1 #2 #3 {\SIrange{#1}{#2}{#3}\xspace}
\DeclareSIUnit\gauss{G}

\newcommand{\myref}[2][]{Fig.~\hyperref[#2]{\ref*{#2}#1}}
\newcommand{\Myref}[2][]{Figure~\hyperref[#2]{\ref*{#2}#1}}
\newcommand{\Mytabref}[2][]{Table~\hyperref[#2]{\ref*{#2}#1}}

\usepackage{bm}
\usepackage{comment}
\usepackage{float} % For figure placement options.
\usepackage[version=4]{mhchem} % Chemical formulae.

\definecolor{orcidlogocol}{HTML}{A6CE39}
\tikzset{
  orcidlogo/.pic={
    \fill[orcidlogocol] svg{M256,128c0,70.7-57.3,128-128,128C57.3,256,0,198.7,0,128C0,57.3,57.3,0,128,0C198.7,0,256,57.3,256,128z};
    \fill[white] svg{M86.3,186.2H70.9V79.1h15.4v48.4V186.2z}
                 svg{M108.9,79.1h41.6c39.6,0,57,28.3,57,53.6c0,27.5-21.5,53.6-56.8,53.6h-41.8V79.1z M124.3,172.4h24.5c34.9,0,42.9-26.5,42.9-39.7c0-21.5-13.7-39.7-43.7-39.7h-23.7V172.4z}
                 svg{M88.7,56.8c0,5.5-4.5,10.1-10.1,10.1c-5.6,0-10.1-4.6-10.1-10.1c0-5.6,4.5-10.1,10.1-10.1C84.2,46.7,88.7,51.3,88.7,56.8z};
  }
}

\newcommand\orcidicon[1]{\href{https://orcid.org/#1}{\mbox{\scalerel*{
\begin{tikzpicture}[yscale=-1,transform shape]
\pic{orcidlogo};
\end{tikzpicture}
}{|}}}}

\begin{document}
\preprint{}

\title{Stability of many-body localization in two dimensions}

\date{\today}

\newcommand{\kaist}[0]{\affiliation{Department of Physics, Korea Advanced Institute of Science and Technology, Daehak 291, 34141, Daejeon, Republic of Korea}}

\newcommand{\tcm}[0]{\affiliation{Theory of Condensed Matter (TCM) Group, Cavendish Laboratory, Department of Physics, J J Thomson Avenue, Cambridge CB3 0HE, United Kingdom}}

\newcommand{\uman}[0]{\affiliation{Department of Physics and Astronomy, University of Manchester, Oxford Road, Manchester M13 9PL, United Kingdom}}

\newcommand{\uw}[0]{\affiliation{Department of Physics, University of Warwick, Coventry, CV4 7AL, United Kingdom}}

\newcommand{\ui}[0]{\affiliation{Universität Innsbruck, Institut für Theoretische Physik, Technikerstraße 21a, 6020 Innsbruck, Austria}}

\newcommand{\iqoqi}[0]{\affiliation{ Institut für Quantenoptik und Quanteninformation, Österreichische Akademie der Wissenschaften, Technikerstraße 21a, 6020 Innsbruck, Austria}}

%\author{ErDy+}
%\affiliation{
 %  Institut f\"{u}r Quantenoptik und Quanteninformation, \"Osterreichische Akademie der Wissenschaften, Innsbruck, Austria
%}
%\affiliation{
 %   Institut f\"{u}r Experimentalphysik, Universit\"{a}t Innsbruck, Austria
%}
 %\affiliation{Department of Physics, University of the Basque Country UPV/EHU, 48080 Bilbao, Spain}
 %\affiliation{IKERBASQUE, Basque Foundation for Science, 48013 Bilbao, Spain}
 
%\author{D.\,S.\,Gr{\"u}n\,\orcidicon{0000-0002-5774-4284}}
%\thanks{These authors contributed equally to this work.}

\author{Junhyeok\, Hur\,\orcidicon{0000-0002-2656-7019}}
\kaist

\author{Joey\, Li\, \orcidicon{0000-0002-8006-2417}}
\ui
\iqoqi

\author{Byungjin\, Lee\,\orcidicon{0009-0006-9177-9057}}
\kaist

\author{Kiryang\, Kwon\,\orcidicon{0000-0002-8252-8436}}
\kaist

\author{Minseok\, Kim\,\orcidicon{0009-0005-1665-170X}}
\kaist

\author{Samgyu\, Hwang\,\orcidicon{0009-0006-3077-288X}}
\kaist

\author{Sumin\, Kim\,\orcidicon{0009-0009-7421-2544}}
\kaist

\author{Yong~\,Soo\, Yu\,\orcidicon{0009-0002-2915-804X}}
\kaist

\author{Amos\,Chan\,\orcidicon{0000-0002-6140-440X}}
\uman
\uw

\author{Thorsten\,Wahl\,\orcidicon{0000-0002-3195-5075}}
\tcm
    
\author{Jae-yoon\,Choi\,\orcidicon{0000-0003-1208-3395}}
\thanks{Email:  \mbox{\url{jaeyoon.choi@kaist.ac.kr}}}
\kaist

%\date{\today}

\begin{abstract}
Disordered quantum many-body systems pose one of the central challenges in condensed matter physics and quantum information science, as their dynamics are generally intractable for classical computation.
Many-body localization (MBL), hypothesized to evade thermalization indefinitely under strong disorder, exemplifies this difficulty.
Here, we study the stability of MBL in two dimensions using ultracold atoms in optical lattices with variable system sizes up to $24\times 24$ sites, well beyond the classically simulable regime.
Using the imbalance as a probe, we trace the long-time dynamics under two distinctive disorder potentials: quasiperiodic and random disorder. For random disorder, the MBL crossover point shifts to higher disorder strength with increasing system size, consistent with the avalanche scenario. In contrast, with quasiperiodic disorder, we observe no clear system size dependence, suggesting possible stability of MBL in two dimensions.
\end{abstract}

\maketitle

%%%%%%%%%%%%%%%%%%%%%%%%%%%%%%%%%%%%%%%%%%%%%%%%%%%%%%%%%%%%%%%%%%%%%%%
%%%%%%%%%%%%%%%%%%%%%%%%%%%%%%%%%%%%%%%%%%%%%%%%%%%%%%%%%%%%%%%%%%%%%%%
%              Introduction:            %
%%%%%%%%%%%%%%%%%%%%%%%%%%%%%%%%%%%%%%%%%%%%%%%%%%%%%%%%%%%%%%%%%%%%%%%
%%%%%%%%%%%%%%%%%%%%%%%%%%%%%%%%%%%%%%%%%%%%%%%%%%%%%%%%%%%%%%%%%%%%%%%

\section*{Introduction}
Statistical physics successfully links the microscopic motion of numerous particles to macroscopic observables. This connection relies on the ergodicity hypothesis, which assumes that all microstates are equally probable. 
In quantum systems, the concept of classical thermalization is reformulated by the eigenstate thermalization hypothesis (ETH)~\cite{Deutsch1991,Srednicki1994,Rigol2008}, according to which eigenstates are locally in Gibbs states. 
For quantum systems with many-body interactions, a naive expectation is that the system will obey ETH, as many-body quantum interference induces ergodicity~\cite{Polkovnikov2011,Eisert2015}. 
Many-body localization (MBL), which is expected to occur in strongly disordered systems, became known as a striking counter-example of this expectation~\cite{Nandkishore2015,Altman2018,Abanin2019}, defying ETH even in the presence of many-body interactions. 
This emergent, non-ergodic phase is of fundamental interest not only for the breakdown of quantum thermalization but also as a potential application for protecting quantum information. 
Experimental studies of MBL, both in one and two dimensions, have been reported in a variety of experimental platforms, including ultracold atoms in optical lattices~\cite{Schreiber2015,Choi2016,Bordia2017,Luschen2017,Luschen2017b,Lukin2019,Rispoli2019,Leonard2023}, trapped ions~\cite{Smith2016}, NV-centers~\cite{Choi2017}, and superconducting qubit arrays~\cite{Roushan2017,Xu2018,Guo2019,Yao2023}. These experiments confirmed key features of the MBL phase, like logarithmic growth of entanglement entropy~\cite{Lukin2019} and level statistics change across the MBL phase transition~\cite{Roushan2017}.

Despite these findings, the existence of the MBL phase under random disorder remains controversial~\cite{DeRoeck2017,Gopalakrishnan2019,Suntajs2020,Sels2021,Peacock2023}.
For example, MBL can be unstable in a random potential because of the thermal avalanche scenario: rare and locally disorder-free regions can act as self-contained baths, gradually melting neighboring localized regions, which ultimately leads to full thermalization of the system in the thermodynamic limit~\cite{DeRoeck2017,Gopalakrishnan2019}. 
On the other hand, such thermal regions may occur so sparsely at strong disorder that they fail to connect into a system-spanning thermal cluster under certain conditions.
In this view, the absence of a percolating network of inclusions prevents avalanches from destabilizing the entire system, allowing MBL to remain stable~\cite{Prelovsek2021}.

\begin{figure*}[t]
\centerline{\includegraphics[width=0.85 \linewidth]{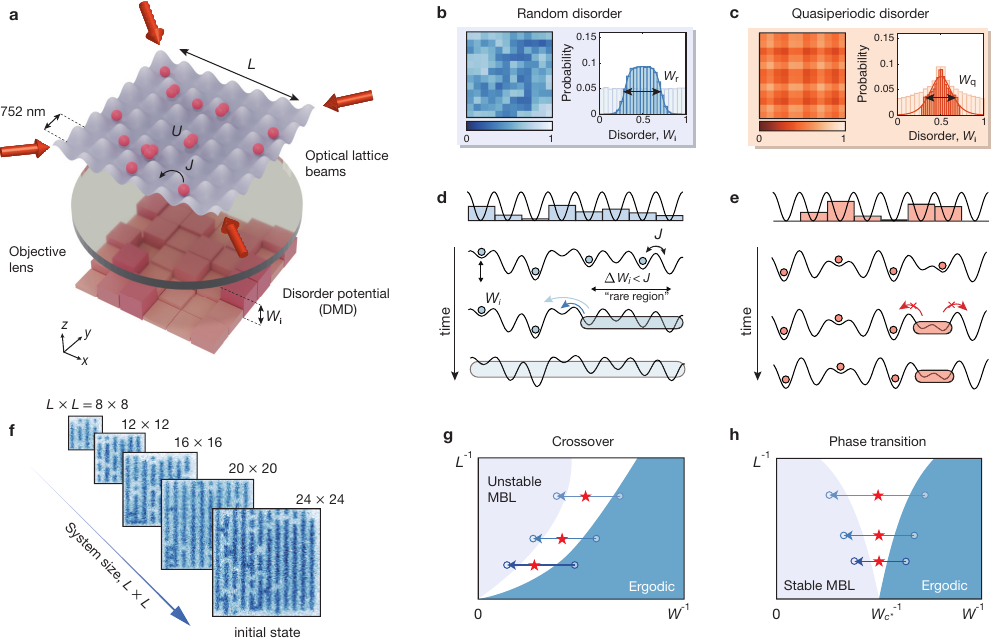}}
\caption{
\textbf{Studying stability of many-body localization with quantum gas microscope.}
\textbf{a,} Illustration of experimental setup. 
Atoms (red spheres) are trapped in an optical lattice with lattice constant $a_{\rm lat}=752\,\text{nm}$, subjected to an external disorder potential generated by a digital micromirror device (DMD). 
The additional strong potential wall generated by the DMD restricts the system size to $L{}\times{}L$. 
During the experiment, two types of disorder are employed: \textbf{b,} uniform random and \textbf{c,} quasiperiodic. 
The comparison between the actual and ideal disorder distributions is shown on the right, with widths labeled $W_{\rm{}r}$ for uniform random and $W_{\rm q}$ for quasiperiodic disorder. 
\textbf{d,} For uniform random disorder, ``rare regions" with locally weak disorder strength ($\Delta W_i<J$) can arise, leading to local thermalization and entire-system thermalization (light blue shading). 
\textbf{e,} In contrast, quasiperiodic disorder does not exhibit large rare regions. 
Thermalization remains confined to small neighboring sites, restricting it to local regions. 
As a result, the system can be stable against this avalanche thermalization mechanism.
\textbf{f,} Representative images of the initial state with various system sizes ($L{}\times{}L$) from $8{}\times {}8$ up to $24\times 24$. 
\textbf{g, h,} Two possible scenarios of the MBL phase. 
For an unstable quantum phase, the ergodic-to-MBL transition point $W_c(L)$ diverges as system size increases, while in the true MBL phase, the transition point converges to $W_{c*}$.
} 
\label{mag-fig1}
\end{figure*}

The stability of the MBL phase becomes even more unclear in higher dimensions. This is the case even in two dimensions (2D), the subject of this work.  While experimental work on optical lattices~\cite{Choi2016,Rubio2019} and supporting numerical simulations \cite{Wahl2019,Chertkov2021} suggest the existence of MBL in 2D, other numerical studies insist on the non-existence of the MBL phase~\cite{Doggen2020}. 
It has been argued that the non-ergodic behavior in the experiments is a transient phenomenon in a finite time and will disappear in the thermodynamic limit.
This ongoing debate is partly driven by the severe challenges in numerically simulating MBL, where the exponentially increasing complexity with system size forces state-of-the-art methods to rely on approximations that can fail in strongly interacting, disordered regimes. 
Consequently, numerical studies of 2D MBL are currently limited to systems no larger than $10 \times 10$ sites~\cite{Wahl2019,Li2024,Doggen2020,Li2025}.
Nevertheless, recent theories suggest that MBL in 2D can be stable in quasiperiodic potentials~\cite{Agrawal2022,Strkalj2022,Crowley2022}, where thermal avalanches can be suppressed because of the absence of rare regions.
Since the rare region effect of a random disorder potential is more prominent in higher dimensions~\cite{DeRoeck2017,Potirniche2019}, a direct comparison of how these two disorder types behave in 2D can address the long-standing question about the stability of MBL.

Here we experimentally study the stability of the two-dimensional MBL phase under two different disorder potentials with ultracold atoms in optical lattices. 
Developing a potential engineering technique in a single-site resolution, we systematically change the system size from $8\times8$  to $24\times24$ lattice sites and create the random disorder and quasiperiodic potential independently. 
This allows for a finite-size analysis of the thermal-to-MBL crossover as well as a direct comparison of disorder types in a single platform. 
For random uniform disorder, the crossover shifts to larger disorder strengths as the system size increases~\cite{Gopalakrishnan2019}, which is inconsistent with a stable two-dimensional MBL phase. In contrast, for quasiperiodic disorder, the crossover is insensitive to system size over the accessible range. These measurements suggest that quasiperiodic disorder is robust to the avalanche instability~\cite{Agrawal2022,Strkalj2022,Crowley2022}. These observations are made possible by reaching large system sizes up to $24 \times 24 = 576$ sites and very long measurement times up to $1500$ hoppings, well beyond the reach of current numerical methods. For example, the numerical simulations in this work (see below and Supplementary Material) required a computation time of approximately five days for each disorder realization for 100 hopping times on an $8 \times 8$ lattice. For $24 \times 24$, the half-system entanglement entropy near the transition would be nine times as large, increasing the required bond dimension by at least the same factor and computational time by $9^3$~\cite{tdvp} to $\sim 10$ years.

%%%%%%%%%%%%%%%%%%%%%%%%%%%%%%%%%%%%%%%%%%%%%%%%%%%%%%%%%%%%%%%%%%%%%
%%%%%%%%%%%%%%%%%%%%%%%%%%%%%%%%%%%%%%%%%%%%%%%%%%%%%%%%%%%%%%%%%%%%%
%              Section 2:    %
%%%%%%%%%%%%%%%%%%%%%%%%%%%%%%%%%%%%%%%%%%%%%%%%%%%%%%%%%%%%%%%%%%%%%

\begin{figure*}[t]
\centerline{\includegraphics[width=0.9\linewidth]{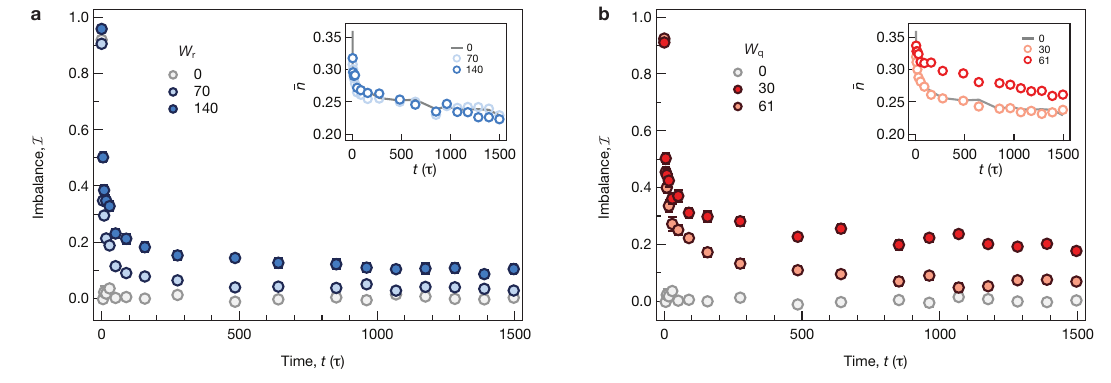}}
\caption{ 
\textbf{Dynamics of imbalance for a $24 \times 24$ system.}    
\textbf{a,} The time-evolution of the imbalance $\mathcal{I}(t)$ under a random disorder potential. 
For sufficiently large disorder, it decreases very slowly and remains finite at long evolution times ($t = 1500\tau$), which signals many-body localization (MBL). 
For a clean system (gray symbols), the imbalance drops to zero, indicating thermalization. 
\textbf{b,} For quasiperiodic disorder, the imbalance remains finite even at weaker disorder strengths, unlike random disorder, where a larger disorder is needed to achieve the localization. This highlights the difference in the dynamics under the two distinct types of disorder. 
Insets show the parity-projected mean density filling $\bar{n}$ under various disorder strengths. 
Similar to the imbalance, the mean density drops rapidly at short times due to doublon formation and then remains almost constant during the last part of the time. The data represented here is obtained by averaging 30-80 iterations.
The error bars, which are smaller than the marker size, denote the one standard error of the mean (s.e.m.).
} \label{mag-fig2}
\end{figure*}

\section*{Experimental Setup}
In our experiment, we investigate 2D MBL using ultracold atoms in optical lattices.
This platform is highly scalable in system size and well isolated from the environment, enabling long-time dynamics and making it ideal for probing localization behavior~\cite{Schreiber2015,Choi2016,Rubio2019,Rispoli2019,Lukin2019,Billy2008,Roati2008,Kondov2011,Yu2024}.
Specifically, our system is described by the two-dimensional disordered Bose-Hubbard Hamiltonian
\begin{align}\label{eq:ham}
H = -J \sum_{\langle {\bf i,j} \rangle} \left( \hat{b}_{\bf i}^\dagger \hat{b}_{\bf j} + \text{h.c.} \right)
+ \frac{U}{2} \sum_{\bf i} \hat{n}_{\bf i} (\hat{n}_{\bf i} - 1) 
+ \sum_{\bf i}  W_{\bf i} \hat{n}_{\bf i}
\end{align}
where $\hat{b}_{\bf i}$ ($\hat{b}_{\bf i}^{\dagger}$) is the bosonic annihilation (creation) operator at site ${\bf i}=(x,y)$, $J$ is the tunneling amplitude, $U$ is the on-site interaction energy, and $W_{\bf i}$ is the random or quasiperiodic disorder potential. 
The tunneling is restricted %exclusively
to nearest-neighbor sites, as indicated by $\langle {\bf i},{\bf j}\rangle$.  

The disorder is generated from a digital micromirror device (DMD), where we project an image generated from the DMD to an atomic plane by using a high-resolution imaging system~\cite{Fukuhara2013,Mazurenko2017}.
The optical projection convolutes the disorder potentials, where we characterize the disorder potentials from atomic responses from the local DMD potential (Fig.~1a and see Supplemental Material). 
The disorder strength is represented by the full width half maximum  (FWHM) of the disorder distribution, $W_\text{r}$ and $W_\text{q}$, for random and quasiperiodic potential, respectively (Fig.~1b and c).
The disorder potential on the DMD is set as follows: in the random case, $W_{\bf i} \in [0, W_\text{r}]$ is sampled from a uniform distribution, and in the quasiperiodic case, $W_{\bf i} =\frac{W_\text{q}}{4}\left[\cos(2\pi\beta_x x + \phi_x)+\cos(2\pi\beta_y y + \phi_y)\right],$ where $\phi_x$ and $\phi_y$ are realization-dependent random phase offsets, and $\beta_x=0.721,\beta_y=0.693$.

\begin{figure*}[t]
\centerline{\includegraphics[width=0.95\linewidth]{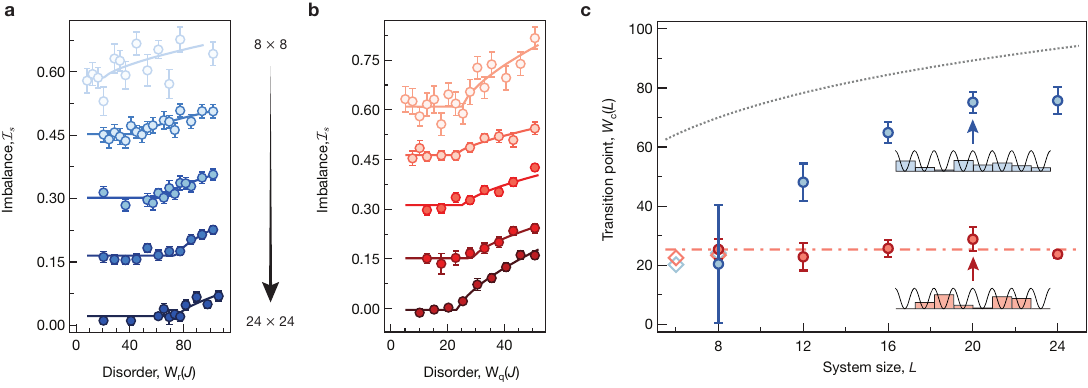}}
\caption{
\textbf{System size scaling of the localization transition.}
\textbf{a,b,} Averaged imbalance for uniform random disorder (\textbf{a}) and quasiperiodic disorder (\textbf{b}) near the critical region, measured across various system sizes from top to bottom, $L\times L=8\times8, 12\times12, 16\times16, 20\times20, {\rm and}~24\times24$. A global offsets of 0.15 between different system sizes are added to the imbalance values for visual clarity. The localization transition points ($W_{c}(L)$) are extracted by fitting the data with an empirical curve (solid line). Error bars represent the standard error of the mean (s.e.m.). Each data point is an average over 50--212 realizations.
\textbf{c,} Extracted localization transition points from system size scaling. The transition points for uniform random disorder (blue symbols) and quasiperiodic disorder (red symbols) are shown.
The gray dashed line represents the asymptotic function $W^{\mathrm{th}}_{c,\mathrm{r}}(L) \sim \text{exp}\left(c_1\ln^{1/3}L^2\right)$ derived from the avalanche instability scenario  as discussed in Ref.~\cite{Gopalakrishnan2019}. 
The red dashed-dot line is a guide to the eye for quasiperiodic disorder. 
Error bars indicate the $1\sigma$ confidence interval obtained from the empirical fit function. Open diamond symbols show transition points for both random and quasiperiodic disorder extracted from numerical simulations (Supplemental Material), for $L=6$ and $L=8$: $W_{c,\textrm{r}}(L = 6) = 20.2 \pm 4.5 J$, $W_{c,\textrm{r}}(L = 8) = 24.1 \pm 3.6 J$, $W_{c,\textrm{q}}(L = 6) = 22.5 \pm 0.7 J$, and $W_{c,\textrm{q}}(L = 8) = 23.4 \pm 1.2 J$.
} \label{mag-fig3}
\end{figure*}

The experiment begins with preparing ultracold Bosonic $^{7}\text{Li}$ atoms in a 2D square optical lattice, where a quantum gas microscope is adopted to have a single site-resolved imaging~\cite{Kwon2022}. 
We take a stripe pattern as an initial state for studying the MBL, where the total system size is controlled by imposing an optical barrier that confines the atoms within $L\times{}L$ sites during the dynamics~(Supplementary Information).
For a given system size $L$, we perform the two independent experiments under the random and quasi-periodic disorder to investigate the dynamical dependence of the localized phase on the disorder types (Fig.~\ref{mag-fig1}d and e).
Even if limited to finite systems, studying the system size dependence of the localized phase is essential for understanding 2D MBL in the thermodynamic limit (Fig.~\ref{mag-fig1}g and h).  
For each type of disorder, we initiate the dynamics by ramping down the optical lattice, turning on the hopping process with interactions $U/J=5.7(2)$, and allowing the system to evolve under the disorder potential. 
Here, the atoms in the optical lattice evolve with the tunneling energy $J/h= 84(2)~\text{Hz}$ with hopping timescale $\tau=\hbar/J=1.89~\text{ms}$.
The strength of the disorder is controlled by changing the incident beam intensity shining onto the DMD. 

%%%%%%%%%%%%%%%%%%%%%%%%%%%%%%%%%%%%%%%%%%%%%%%%%%%%%%%%%%%%%%%%%%%%%%%
%%%%%%%%%%%%%%%%%%%%%%%%%%%%%%%%%%%%%%%%%%%%%%%%%%%%%%%%%%%%%%%%%%%%%%%
%              Fig 2: Time-evolution of 2D MBL   %
%%%%%%%%%%%%%%%%%%%%%%%%%%%%%%%%%%%%%%%%%%%%%%%%%%%%%%%%%%%%%%%%%%%%%%%
%%%%%%%%%%%%%%%%%%%%%%%%%%%%%%%%%%%%%%%%%%%%%%%%%%%%%%%%%%%%%%%%%%%%

\section*{Breakdown of thermalization}
Following the time evolution, we measured the atomic distribution and used the imbalance $\mathcal{I}=(n_{\rm odd} - n_{\rm  even})/(n_{\rm odd} + n_{\rm even})$ as an indicator for localization. 
Here, $n_{\rm odd}$ ($n_{\rm even}$) represents the parity-projected occupation number of atoms on odd (even) lattice columns. %sites. 
This observable can quantify the amount of initial state memory so that the imbalance shows a finite value in the localized regime~\cite{Schreiber2015,Choi2016,Morong2021}.

Figure~2 displays the time evolution of the imbalance $\mathcal{I}(t)$ at system size $L\times L = 24 \times 24$ under random and quasiperiodic potential with various disorder strengths. 
For both cases, the imbalance shows a clear dependence on the disorder strength. 
In the absence of disorder, the imbalance rapidly drops to zero within a few tunneling times, while under the disorder potential, it decays slowly and eventually saturates to a finite value after a long time evolution.
The dynamics of the imbalance under random disorder potential at $W_\text{r} = 147{J} $ are similar to those in the previous 2D MBL experiment~\cite{Rubio2019}. 
Right after the quench, the imbalance quickly drops due to the tunneling process, and then it gradually decays because of doublon formation. 
This can be seen in the dynamics of the atomic density, which decays on the same time scale as the imbalance.
An exponential fit of the initial dynamics yields {3.6(7)$\tau$ and 3.1(5)$\tau$} for the decay time of the imbalance and density, respectively.

Dynamics under a disorder potential can be slow, and a complete study of the relaxation dynamics might not be accessible in real experiments~\cite{Sierant2025}. 
In our experiment, the mean density shows a long decay time, which depends on the disorder strength.
For example, for the quasiperiodic disorder (Fig.~2b inset), we find the mean density gradually decays over the entire hold time at strong disorder potential,  $W_\text{q}=61J$. Note that the density decay cannot be attributed to atom loss during the time evolution, for example, from the off-resonant scattering of the DMD potential, since its absolute value is higher under a stronger disorder potential. The slow dynamics of the doublons are also reflected in the imbalance (Fig.~2b), where an exponential fit of the imbalance shows the relaxation time to be $t\approx 4000\tau$. When the doublon dynamics reach a quasi steady state (Fig.~2b at $W_\text{q}=30J$), the imbalance also stays at a nearly constant value. Therefore, we call this regime the ``MBL regime", where the relaxation time scale is so slow that we might not be able to discern the true localization within our experimental time window.

%%%%%%%%%%%%%%%%%%%%%%%%%%%%%%%%%%%%%%%%%%%%%%%%%%%%%%%%%%%%%%%%%%%%%%%
%%%%%%%%%%%%%%%%%%%%%%%%%%%%%%%%%%%%%%%%%%%%%%%%%%%%%%%%%%%%%%%%%%%%%%%
%              Fig 3: System size scaling   %
%%%%%%%%%%%%%%%%%%%%%%%%%%%%%%%%%%%%%%%%%%%%%%%%%%%%%%%%%%%%%%%%%%%%%%%
%%%%%%%%%%%%%%%%%%%%%%%%%%%%%%%%%%%%%%%%%%%%%%%%%%%%%%%%%%%%%%%%%%%%%%%

\section*{System size scaling}
Within the experimentally accessible time window, investigating the system size dependence of the MBL transition becomes an effective strategy to assess the stability of MBL in two dimensions. 
By tracking how the transition point shifts with increasing system size, we can test whether the observed MBL is consistent with a stable phase that could persist in the thermodynamic limit, or instead is a prethermal phenomenon that thermalizes at longer times (Fig.~1g and h).
To probe the transition point for a given system size, we set the evolution time to its maximum value of $t_s=1500\tau$ and measured the imbalance at $t_s$, $\mathcal{I}_s=\mathcal{I}(t_s)$, under various disorder strengths. 
As the applied disorder strength increases, the onset of imbalance is observed, signaling the emergence of the ``MBL regime" for both types of disorder.
The transition curves for various system sizes ($L\times L=8\times8, 12\times12, 16\times16, 20\times20, 24\times24$) are plotted in Fig.~\ref{mag-fig3}a (random disorder) and in Fig.~\ref{mag-fig3}b (quasiperiodic disorder), respectively.

To investigate the system size dependence of the finite time transition, we extract the critical disorder through an empirical fit to the transition curve, $\mathcal{I}_s(W_\text{r(q)})=a(W_\text{r(q)}-W_c)^{2/3}+b$ for $W_{\rm r(q)}>W_c$ and $b$ otherwise, where $W_c$ is the critical disorder.
Fig.~\ref{mag-fig3}c displays the extracted critical disorder for both disorder potentials as a function of the system size.
For random disorder, the extracted transition points ($W_{c, r}$) exhibit a clear shift towards higher values as the system size increases. 
In contrast, for quasiperiodic disorder, the transition point ($W_{c,q}$) shows no significant dependence on system size across all measured system sizes up to $24\times{}24=576$ sites.
These results are consistent with numerical simulations (for small system sizes, $L=6,8$) using the time-dependent variational principle (TDVP)~\cite{tdvp} (see Supplemental Material for details).

The distinct system size scaling observed for these two disorder types suggests that the mechanism of thermalization in disordered systems can vary depending on the specific nature of the disorder.
In the avalanche scenario uniform random disorder displays an asymptotic shift of the transition point as $W^{\mathrm{th}}_{c,\mathrm{r}}(L) \sim \text{exp}(c_1\ln^{1/3}L^2)$~\cite{Gopalakrishnan2019} with $c_1 = 2(\ln2)^{2/3}\approx1.57$ \cite{Doggen2020}.
This asymptotic curve, depicted as a dotted line in Fig.~\ref{mag-fig3}, assumes a large system size and is consistent with our experimental data at larger system sizes. 
By following the asymptotic curve, the avalanche instability ultimately breaks the MBL phase in larger dimensions, leading to thermalization.

In contrast, the data for the quasiperiodic disorder distribution does not follow this avalanche asymptotic curve. 
Instead, it aligns with a constant value. 
This difference indicates that these two types of disorder possess intrinsically distinct natures of thermalization with increasing system size. 
Our results support the hypothesis that systems with quasiperiodic disorder in our quantum simulator remain stable against avalanche instability up to sizes of $L \times L = 24 \times 24$.

%%%%%%%%%%%%%%%%%%%%%%%%%%%%%%%%%%%%%%%%%%%%%%%%%%%%%%%%%%%%%%%%%%%%%%%
%%%%%%%%%%%%%%%%%%%%%%%%%%%%%%%%%%%%%%%%%%%%%%%%%%%%%%%%%%%%%%%%%%%%%%%
%              Fig 4: Thermalization of system with time    %
%%%%%%%%%%%%%%%%%%%%%%%%%%%%%%%%%%%%%%%%%%%%%%%%%%%%%%%%%%%%%%%%%%%%%%%
\begin{figure}[t!]
\centerline{\includegraphics[width=0.9 \linewidth]{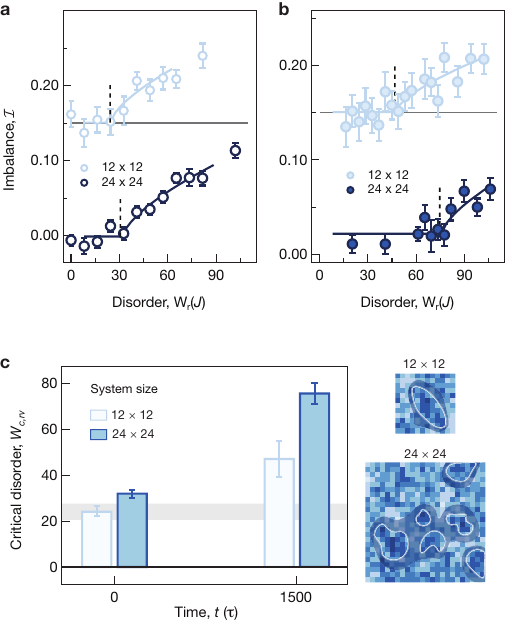}}
\caption{
\textbf{Shift of critical disorder with evolution time.}
\textbf{a,} Imbalance as a function of the random disorder strength at hold time $t=150\tau$ for $L\times L=12 \times 12$ and $24 \times 24$. 
\textbf{b,} Imbalance after a longer time evolution $t=1500\tau$ for the two different system sizes. 
For visual clarity, the data for the $12 \times 12$ system is shifted vertically by an offset of 0.15, marked by the horizontal lines. 
The solid lines represent fit curves to the data, and the vertical dashed line marks the extracted localization transition point, $W_{c,r}$.
Each data point was calculated by averaging 50 to 150 individual measurements.
\textbf{c}, Transition points of random disorder potential for short ($t=150\tau$) and long ($t=1500\tau$) evolution times with two system sizes. 
Gray shading is the critical disorder from the numerical simulation with $L=8$ and ($t=100\tau$). 
The error bars represent 1$\sigma$ confidence bounds of the fit parameters. 
The images on the right side schematically illustrate the avalanche mechanism from thermal bubbles (white lines). 
At short evolution times, the size of these bubbles is small and isolated, while in later times they can meet to form bigger bubbles (shaded region). 
} \label{mag-fig4}
\end{figure}

\section*{Quantum avalanche with time}
In the avalanche mechanism for a random disorder potential, the thermalization time can depend on the system size. This is because the rare region with weak disorder can generate local thermal bubbles, which act as thermal baths. After a very long time, these thermal bubbles grow and separate bubbles can merge, then the entire system can eventually thermalize~\cite{DeRoeck2017,Gopalakrishnan2019}.
Specifically, avalanche theory predicts that for fixed system size $L \times L$, an ergodic grain takes time $t_\mathrm{deloc}(L) \sim \exp(\mathrm{const} \cdot L)$ to delocalize the entire system~\cite{Gopalakrishnan2019}. 
This implies that if the observation time $t < t_\mathrm{deloc}(L)$ for all considered $L$, the experimentally measured transition point will only depend on the evolution time, $W_c(t) \sim \exp[\mathrm{const} \cdot (\ln \ln t)^{1/3}]$, and not on the system size.

Following this prediction, we perform the experiments that investigate the time- and size-dependence of the ergodic-to-MBL crossover in random potential and plot the results in Fig.~\ref{mag-fig4}.
The characteristic transition point at short time ($t=150\tau$) is slightly increased with larger system size (Fig.~\ref{mag-fig4}a), and the critical disorder with $12\times12$ systems is similar to the numerically obtained critical point.  
In contrast, at a later time ($t=1500\tau$), noticeable system size dependence is observed in the transition point (Fig.~\ref{mag-fig4}b and c). 
Our observation is consistent with the avalanche theory, where at short times the size of thermal bubbles is not sufficiently large to fully thermalize the system (Fig.~\ref{mag-fig4}c inset), such that the critical disorder dependence on the system size becomes weak.

%%%%%%%%%%%%%%%%%%%%%%%%%%%%%%%%%%%%%%%%%%%%%%%%%%%%%%%%%%%%%%%%%%%%%%%
%%%%%%%%%%%%%%%%%%%%%%%%%%%%%%%%%%%%%%%%%%%%%%%%%%%%%%%%%%%%%%%%%%%%%%%
%              Conclusion              %
%%%%%%%%%%%%%%%%%%%%%%%%%%%%%%%%%%%%%%%%%%%%%%%%%%%%%%%%%%%%%%%%%%%%%%%

\section*{Conclusions and outlook}
We performed experiments on the stability of two-dimensional many-body localization (MBL), reaching system sizes well beyond the numerically accessible. Our results provide experimental evidence that random-disorder MBL in 2D is fragile to avalanche instability, by showing the upward drift of the transition point as a function of system size. However, we find that MBL induced by quasiperiodic disorder remains stable. Taken together, these observations support a disorder-dependent scenario: true random disorder allows for the appearance of rare local thermal regions that create avalanches that restore ergodicity, while quasiperiodicity suppresses such rare regions. Further work might include a more detailed experimental probe of the avalanche phenomenon--for instance, one might engineer a controlled thermal bubble, i.e., a locally weakened disorder patch, and track its growth through space and time~\cite{Chandran2016}. These studies could also test the stability of the 2D MBL phase in quasiperiodic potential~\cite{Agrawal2022,Strkalj2022,Crowley2022}. In particular, an intriguing coexistence of localization and transport can be studied in the Aubry–André model, where particle transport occurs along deterministic weak potential lines~\cite{Strkalj2022}. Moreover, our platform enables systematic finite-size scaling in two dimensions, which allows quantitative studies of critical phenomena at the ergodic-to-MBL phase transition in quasiperiodic models~\cite{Khemani2017,Agrawal2020}.\\

\textit{Note added:} While completing this manuscript, we learned of a recent complementary quantum simulation on a 70-qubit superconducting processor for uncorrelated random disorder~\cite{70qubit2dMBL}.\\

\begin{acknowledgments}
We acknowledge discussions with Soonwon Choi, Joonhee Choi, David. A. Huse, Johannes Zeiher, Seung-Jung Huh, Kyungtae Kim, and Elmer Doggen. 
J.-y.C. is supported by the Samsung Science and Technology Foundation (Grant No. BA1702-06) and the National Research Foundation of Korea (NRF) Grant under Project No. RS-2023-00207974, RS-2023-00218998, and 2023M3K5A1094812.
T.B.W. is grateful for support from an EPSRC ERC underwrite Grant No. EP/X025829/1. J.L. acknowledges support from ERC Starting grant QARA (Grant No.~101041435) and the Austrian Science Fund (FWF): COE 1 and quantA. A.C. acknowledges support from the EPSRC Open Fellowship EP/X042812/1. The computational results presented here have been achieved (in part) using the LEO HPC infrastructure of the University of Innsbruck.

\end{acknowledgments}

\bibliography{Ref_MBL2D}

\begin{thebibliography}{59}%
\makeatletter
\providecommand \@ifxundefined [1]{%
 \@ifx{#1\undefined}
}%
\providecommand \@ifnum [1]{%
 \ifnum #1\expandafter \@firstoftwo
 \else \expandafter \@secondoftwo
 \fi
}%
\providecommand \@ifx [1]{%
 \ifx #1\expandafter \@firstoftwo
 \else \expandafter \@secondoftwo
 \fi
}%
\providecommand \natexlab [1]{#1}%
\providecommand \enquote  [1]{``#1''}%
\providecommand \bibnamefont  [1]{#1}%
\providecommand \bibfnamefont [1]{#1}%
\providecommand \citenamefont [1]{#1}%
\providecommand \href@noop [0]{\@secondoftwo}%
\providecommand \href [0]{\begingroup \@sanitize@url \@href}%
\providecommand \@href[1]{\@@startlink{#1}\@@href}%
\providecommand \@@href[1]{\endgroup#1\@@endlink}%
\providecommand \@sanitize@url [0]{\catcode `\\12\catcode `\$12\catcode
  `\&12\catcode `\#12\catcode `\^12\catcode `\_12\catcode `\%12\relax}%
\providecommand \@@startlink[1]{}%
\providecommand \@@endlink[0]{}%
\providecommand \url  [0]{\begingroup\@sanitize@url \@url }%
\providecommand \@url [1]{\endgroup\@href {#1}{\urlprefix }}%
\providecommand \urlprefix  [0]{URL }%
\providecommand \Eprint [0]{\href }%
\providecommand \doibase [0]{https://doi.org/}%
\providecommand \selectlanguage [0]{\@gobble}%
\providecommand \bibinfo  [0]{\@secondoftwo}%
\providecommand \bibfield  [0]{\@secondoftwo}%
\providecommand \translation [1]{[#1]}%
\providecommand \BibitemOpen [0]{}%
\providecommand \bibitemStop [0]{}%
\providecommand \bibitemNoStop [0]{.\EOS\space}%
\providecommand \EOS [0]{\spacefactor3000\relax}%
\providecommand \BibitemShut  [1]{\csname bibitem#1\endcsname}%
\let\auto@bib@innerbib\@empty
%</preamble>
\bibitem [{\citenamefont {Deutsch}(1991)}]{Deutsch1991}%
  \BibitemOpen
  \bibfield  {author} {\bibinfo {author} {\bibfnamefont {J.~M.}\ \bibnamefont
  {Deutsch}},\ }\bibfield  {title} {\bibinfo {title} {{Quantum statistical
  mechanics in a closed system}},\ }\href
  {https://doi.org/10.1103/PhysRevA.43.2046} {\bibfield  {journal} {\bibinfo
  {journal} {Phys. Rev. A}\ }\textbf {\bibinfo {volume} {43}},\ \bibinfo
  {pages} {2046–2049} (\bibinfo {year} {1991})}\BibitemShut {NoStop}%
\bibitem [{\citenamefont {Srednicki}(1994)}]{Srednicki1994}%
  \BibitemOpen
  \bibfield  {author} {\bibinfo {author} {\bibfnamefont {M.}~\bibnamefont
  {Srednicki}},\ }\bibfield  {title} {\bibinfo {title} {{Chaos and quantum
  thermalization}},\ }\href {https://doi.org/10.1103/PhysRevE.50.888}
  {\bibfield  {journal} {\bibinfo  {journal} {Phys. Rev. E}\ }\textbf {\bibinfo
  {volume} {50}},\ \bibinfo {pages} {888–901} (\bibinfo {year}
  {1994})}\BibitemShut {NoStop}%
\bibitem [{\citenamefont {Rigol}\ \emph {et~al.}(2008)\citenamefont {Rigol},
  \citenamefont {Dunjko},\ and\ \citenamefont {Olshanii}}]{Rigol2008}%
  \BibitemOpen
  \bibfield  {author} {\bibinfo {author} {\bibfnamefont {M.}~\bibnamefont
  {Rigol}}, \bibinfo {author} {\bibfnamefont {V.}~\bibnamefont {Dunjko}},\ and\
  \bibinfo {author} {\bibfnamefont {M.}~\bibnamefont {Olshanii}},\ }\bibfield
  {title} {\bibinfo {title} {{Thermalization and its mechanism for generic
  isolated quantum systems}},\ }\href {https://doi.org/10.1038/nature06838}
  {\bibfield  {journal} {\bibinfo  {journal} {Nature}\ }\textbf {\bibinfo
  {volume} {452}},\ \bibinfo {pages} {854–858} (\bibinfo {year}
  {2008})}\BibitemShut {NoStop}%
\bibitem [{\citenamefont {Polkovnikov}\ \emph {et~al.}(2011)\citenamefont
  {Polkovnikov}, \citenamefont {Sengupta}, \citenamefont {Silva},\ and\
  \citenamefont {Vengalattore}}]{Polkovnikov2011}%
  \BibitemOpen
  \bibfield  {author} {\bibinfo {author} {\bibfnamefont {A.}~\bibnamefont
  {Polkovnikov}}, \bibinfo {author} {\bibfnamefont {K.}~\bibnamefont
  {Sengupta}}, \bibinfo {author} {\bibfnamefont {A.}~\bibnamefont {Silva}},\
  and\ \bibinfo {author} {\bibfnamefont {M.}~\bibnamefont {Vengalattore}},\
  }\bibfield  {title} {\bibinfo {title} {{Colloquium: Nonequilibrium dynamics
  of closed interacting quantum systems}},\ }\href
  {https://doi.org/10.1103/RevModPhys.83.863} {\bibfield  {journal} {\bibinfo
  {journal} {Rev. Mod. Phys.}\ }\textbf {\bibinfo {volume} {83}},\ \bibinfo
  {pages} {863–883} (\bibinfo {year} {2011})}\BibitemShut {NoStop}%
\bibitem [{\citenamefont {Eisert}\ \emph {et~al.}(2015)\citenamefont {Eisert},
  \citenamefont {Friesdorf},\ and\ \citenamefont {Gogolin}}]{Eisert2015}%
  \BibitemOpen
  \bibfield  {author} {\bibinfo {author} {\bibfnamefont {J.}~\bibnamefont
  {Eisert}}, \bibinfo {author} {\bibfnamefont {M.}~\bibnamefont {Friesdorf}},\
  and\ \bibinfo {author} {\bibfnamefont {C.}~\bibnamefont {Gogolin}},\
  }\bibfield  {title} {\bibinfo {title} {{Quantum many-body systems out of
  equilibrium}},\ }\href {https://doi.org/10.1038/nphys3215} {\bibfield
  {journal} {\bibinfo  {journal} {Nat. Phys.}\ }\textbf {\bibinfo {volume}
  {11}},\ \bibinfo {pages} {124–130} (\bibinfo {year} {2015})}\BibitemShut
  {NoStop}%
\bibitem [{\citenamefont {Nandkishore}\ and\ \citenamefont
  {Huse}(2015)}]{Nandkishore2015}%
  \BibitemOpen
  \bibfield  {author} {\bibinfo {author} {\bibfnamefont {R.}~\bibnamefont
  {Nandkishore}}\ and\ \bibinfo {author} {\bibfnamefont {D.~A.}\ \bibnamefont
  {Huse}},\ }\bibfield  {title} {\bibinfo {title} {{Many-Body Localization and
  Thermalization in Quantum Statistical Mechanics}},\ }\href
  {https://doi.org/10.1146/annurev-conmatphys-031214-014726} {\bibfield
  {journal} {\bibinfo  {journal} {Annu. Rev. Condens. Matter Phys.}\ }\textbf
  {\bibinfo {volume} {6}},\ \bibinfo {pages} {15–38} (\bibinfo {year}
  {2015})}\BibitemShut {NoStop}%
\bibitem [{\citenamefont {Altman}(2018)}]{Altman2018}%
  \BibitemOpen
  \bibfield  {author} {\bibinfo {author} {\bibfnamefont {E.}~\bibnamefont
  {Altman}},\ }\bibfield  {title} {\bibinfo {title} {{ Many-body localization
  and quantum thermalization}},\ }\href
  {https://doi.org/10.1038/s41567-018-0305-7} {\bibfield  {journal} {\bibinfo
  {journal} {Nat. Phys.}\ }\textbf {\bibinfo {volume} {14}},\ \bibinfo {pages}
  {979–983} (\bibinfo {year} {2018})}\BibitemShut {NoStop}%
\bibitem [{\citenamefont {Abanin}\ \emph {et~al.}(2019)\citenamefont {Abanin},
  \citenamefont {Altman}, \citenamefont {Bloch},\ and\ \citenamefont
  {Serbyn}}]{Abanin2019}%
  \BibitemOpen
  \bibfield  {author} {\bibinfo {author} {\bibfnamefont {D.~A.}\ \bibnamefont
  {Abanin}}, \bibinfo {author} {\bibfnamefont {E.}~\bibnamefont {Altman}},
  \bibinfo {author} {\bibfnamefont {I.}~\bibnamefont {Bloch}},\ and\ \bibinfo
  {author} {\bibfnamefont {M.}~\bibnamefont {Serbyn}},\ }\bibfield  {title}
  {\bibinfo {title} {{Colloquium: Many-body localization, thermalization, and
  entanglement}},\ }\href {https://doi.org/10.1103/RevModPhys.91.021001}
  {\bibfield  {journal} {\bibinfo  {journal} {Rev. Mod. Phys.}\ }\textbf
  {\bibinfo {volume} {91}},\ \bibinfo {pages} {021001} (\bibinfo {year}
  {2019})}\BibitemShut {NoStop}%
\bibitem [{\citenamefont {Schreiber}\ \emph {et~al.}(2015)\citenamefont
  {Schreiber}, \citenamefont {Hodgman}, \citenamefont {Bordia}, \citenamefont
  {L{\"u}schen}, \citenamefont {Fischer}, \citenamefont {Vosk}, \citenamefont
  {Altman}, \citenamefont {Schneider},\ and\ \citenamefont
  {Bloch}}]{Schreiber2015}%
  \BibitemOpen
  \bibfield  {author} {\bibinfo {author} {\bibfnamefont {M.}~\bibnamefont
  {Schreiber}}, \bibinfo {author} {\bibfnamefont {S.~S.}\ \bibnamefont
  {Hodgman}}, \bibinfo {author} {\bibfnamefont {P.}~\bibnamefont {Bordia}},
  \bibinfo {author} {\bibfnamefont {H.~P.}\ \bibnamefont {L{\"u}schen}},
  \bibinfo {author} {\bibfnamefont {M.~H.}\ \bibnamefont {Fischer}}, \bibinfo
  {author} {\bibfnamefont {R.}~\bibnamefont {Vosk}}, \bibinfo {author}
  {\bibfnamefont {E.}~\bibnamefont {Altman}}, \bibinfo {author} {\bibfnamefont
  {U.}~\bibnamefont {Schneider}},\ and\ \bibinfo {author} {\bibfnamefont
  {I.}~\bibnamefont {Bloch}},\ }\bibfield  {title} {\bibinfo {title}
  {{Observation of many-body localization of interacting fermions in a
  quasirandom optical lattice}},\ }\href
  {https://doi.org/10.1126/science.aaa7432} {\bibfield  {journal} {\bibinfo
  {journal} {Science}\ }\textbf {\bibinfo {volume} {349}},\ \bibinfo {pages}
  {842–845} (\bibinfo {year} {2015})}\BibitemShut {NoStop}%
\bibitem [{\citenamefont {Choi}\ \emph {et~al.}(2016)\citenamefont {Choi},
  \citenamefont {Hild}, \citenamefont {Zeiher}, \citenamefont {Schau{\ss}},
  \citenamefont {Rubio-Abadal}, \citenamefont {Yefsah}, \citenamefont
  {Khemani}, \citenamefont {Huse}, \citenamefont {Bloch},\ and\ \citenamefont
  {Gross}}]{Choi2016}%
  \BibitemOpen
  \bibfield  {author} {\bibinfo {author} {\bibfnamefont {J.-y.}\ \bibnamefont
  {Choi}}, \bibinfo {author} {\bibfnamefont {S.}~\bibnamefont {Hild}}, \bibinfo
  {author} {\bibfnamefont {J.}~\bibnamefont {Zeiher}}, \bibinfo {author}
  {\bibfnamefont {P.}~\bibnamefont {Schau{\ss}}}, \bibinfo {author}
  {\bibfnamefont {A.}~\bibnamefont {Rubio-Abadal}}, \bibinfo {author}
  {\bibfnamefont {T.}~\bibnamefont {Yefsah}}, \bibinfo {author} {\bibfnamefont
  {V.}~\bibnamefont {Khemani}}, \bibinfo {author} {\bibfnamefont {D.~A.}\
  \bibnamefont {Huse}}, \bibinfo {author} {\bibfnamefont {I.}~\bibnamefont
  {Bloch}},\ and\ \bibinfo {author} {\bibfnamefont {C.}~\bibnamefont {Gross}},\
  }\bibfield  {title} {\bibinfo {title} {{Exploring the many-body localization
  transition in two dimensions}},\ }\href
  {https://doi.org/10.1126/science.aaf8834} {\bibfield  {journal} {\bibinfo
  {journal} {Science}\ }\textbf {\bibinfo {volume} {352}},\ \bibinfo {pages}
  {1547–1552} (\bibinfo {year} {2016})}\BibitemShut {NoStop}%
\bibitem [{\citenamefont {Bordia}\ \emph {et~al.}(2017)\citenamefont {Bordia},
  \citenamefont {L\"uschen}, \citenamefont {Scherg}, \citenamefont
  {Gopalakrishnan}, \citenamefont {Knap}, \citenamefont {Schneider},\ and\
  \citenamefont {Bloch}}]{Bordia2017}%
  \BibitemOpen
  \bibfield  {author} {\bibinfo {author} {\bibfnamefont {P.}~\bibnamefont
  {Bordia}}, \bibinfo {author} {\bibfnamefont {H.}~\bibnamefont {L\"uschen}},
  \bibinfo {author} {\bibfnamefont {S.}~\bibnamefont {Scherg}}, \bibinfo
  {author} {\bibfnamefont {S.}~\bibnamefont {Gopalakrishnan}}, \bibinfo
  {author} {\bibfnamefont {M.}~\bibnamefont {Knap}}, \bibinfo {author}
  {\bibfnamefont {U.}~\bibnamefont {Schneider}},\ and\ \bibinfo {author}
  {\bibfnamefont {I.}~\bibnamefont {Bloch}},\ }\bibfield  {title} {\bibinfo
  {title} {{Probing Slow Relaxation and Many-Body Localization in
  Two-Dimensional Quasiperiodic Systems}},\ }\href
  {https://doi.org/10.1103/PhysRevX.7.041047} {\bibfield  {journal} {\bibinfo
  {journal} {Phys. Rev. X}\ }\textbf {\bibinfo {volume} {7}},\ \bibinfo {pages}
  {041047} (\bibinfo {year} {2017})}\BibitemShut {NoStop}%
\bibitem [{\citenamefont {Lüschen}\ \emph
  {et~al.}(2017{\natexlab{a}})\citenamefont {Lüschen}, \citenamefont {Bordia},
  \citenamefont {Hodgman}, \citenamefont {Schreiber}, \citenamefont {Sarkar},
  \citenamefont {Daley}, \citenamefont {Fischer}, \citenamefont {Altman},
  \citenamefont {Bloch},\ and\ \citenamefont {Schneider}}]{Luschen2017}%
  \BibitemOpen
  \bibfield  {author} {\bibinfo {author} {\bibfnamefont {H.~P.}\ \bibnamefont
  {Lüschen}}, \bibinfo {author} {\bibfnamefont {P.}~\bibnamefont {Bordia}},
  \bibinfo {author} {\bibfnamefont {S.~S.}\ \bibnamefont {Hodgman}}, \bibinfo
  {author} {\bibfnamefont {M.}~\bibnamefont {Schreiber}}, \bibinfo {author}
  {\bibfnamefont {S.}~\bibnamefont {Sarkar}}, \bibinfo {author} {\bibfnamefont
  {A.~J.}\ \bibnamefont {Daley}}, \bibinfo {author} {\bibfnamefont {M.~H.}\
  \bibnamefont {Fischer}}, \bibinfo {author} {\bibfnamefont {E.}~\bibnamefont
  {Altman}}, \bibinfo {author} {\bibfnamefont {I.}~\bibnamefont {Bloch}},\ and\
  \bibinfo {author} {\bibfnamefont {U.}~\bibnamefont {Schneider}},\ }\bibfield
  {title} {\bibinfo {title} {{Signatures of Many-Body Localization in a
  Controlled Open Quantum System}},\ }\href
  {https://doi.org/10.1103/PhysRevX.7.011034} {\bibfield  {journal} {\bibinfo
  {journal} {Phys. Rev. X}\ }\textbf {\bibinfo {volume} {7}},\ \bibinfo {pages}
  {011034} (\bibinfo {year} {2017}{\natexlab{a}})}\BibitemShut {NoStop}%
\bibitem [{\citenamefont {Lüschen}\ \emph
  {et~al.}(2017{\natexlab{b}})\citenamefont {Lüschen}, \citenamefont {Bordia},
  \citenamefont {Scherg}, \citenamefont {Alet}, \citenamefont {Altman},
  \citenamefont {Schneider},\ and\ \citenamefont {Bloch}}]{Luschen2017b}%
  \BibitemOpen
  \bibfield  {author} {\bibinfo {author} {\bibfnamefont {H.~P.}\ \bibnamefont
  {Lüschen}}, \bibinfo {author} {\bibfnamefont {P.}~\bibnamefont {Bordia}},
  \bibinfo {author} {\bibfnamefont {S.}~\bibnamefont {Scherg}}, \bibinfo
  {author} {\bibfnamefont {F.}~\bibnamefont {Alet}}, \bibinfo {author}
  {\bibfnamefont {E.}~\bibnamefont {Altman}}, \bibinfo {author} {\bibfnamefont
  {U.}~\bibnamefont {Schneider}},\ and\ \bibinfo {author} {\bibfnamefont
  {I.}~\bibnamefont {Bloch}},\ }\bibfield  {title} {\bibinfo {title}
  {{Observation of Slow Dynamics near the Many-Body Localization Transition in
  One-Dimensional Quasiperiodic Systems}},\ }\href
  {https://doi.org/10.1103/PhysRevLett.119.260401} {\bibfield  {journal}
  {\bibinfo  {journal} {Phys. Rev. Lett.}\ }\textbf {\bibinfo {volume} {119}},\
  \bibinfo {pages} {260401} (\bibinfo {year} {2017}{\natexlab{b}})}\BibitemShut
  {NoStop}%
\bibitem [{\citenamefont {Lukin}\ \emph {et~al.}(2019)\citenamefont {Lukin},
  \citenamefont {Rispoli}, \citenamefont {Schittko}, \citenamefont {Tai},
  \citenamefont {Kaufman}, \citenamefont {Choi}, \citenamefont {Khemani},
  \citenamefont {L{\'e}onard},\ and\ \citenamefont {Greiner}}]{Lukin2019}%
  \BibitemOpen
  \bibfield  {author} {\bibinfo {author} {\bibfnamefont {A.}~\bibnamefont
  {Lukin}}, \bibinfo {author} {\bibfnamefont {M.}~\bibnamefont {Rispoli}},
  \bibinfo {author} {\bibfnamefont {R.}~\bibnamefont {Schittko}}, \bibinfo
  {author} {\bibfnamefont {M.~E.}\ \bibnamefont {Tai}}, \bibinfo {author}
  {\bibfnamefont {A.~M.}\ \bibnamefont {Kaufman}}, \bibinfo {author}
  {\bibfnamefont {S.}~\bibnamefont {Choi}}, \bibinfo {author} {\bibfnamefont
  {V.}~\bibnamefont {Khemani}}, \bibinfo {author} {\bibfnamefont
  {J.}~\bibnamefont {L{\'e}onard}},\ and\ \bibinfo {author} {\bibfnamefont
  {M.}~\bibnamefont {Greiner}},\ }\bibfield  {title} {\bibinfo {title}
  {{Probing entanglement in a many-body–localized system}},\ }\href
  {https://doi.org/10.1126/science.aau0818} {\bibfield  {journal} {\bibinfo
  {journal} {Science}\ }\textbf {\bibinfo {volume} {364}},\ \bibinfo {pages}
  {256–260} (\bibinfo {year} {2019})}\BibitemShut {NoStop}%
\bibitem [{\citenamefont {Rispoli}\ \emph {et~al.}(2019)\citenamefont
  {Rispoli}, \citenamefont {Lukin}, \citenamefont {Schittko}, \citenamefont
  {Kim}, \citenamefont {Tai}, \citenamefont {L{\'e}onard},\ and\ \citenamefont
  {Greiner}}]{Rispoli2019}%
  \BibitemOpen
  \bibfield  {author} {\bibinfo {author} {\bibfnamefont {M.}~\bibnamefont
  {Rispoli}}, \bibinfo {author} {\bibfnamefont {A.}~\bibnamefont {Lukin}},
  \bibinfo {author} {\bibfnamefont {R.}~\bibnamefont {Schittko}}, \bibinfo
  {author} {\bibfnamefont {S.}~\bibnamefont {Kim}}, \bibinfo {author}
  {\bibfnamefont {M.~E.}\ \bibnamefont {Tai}}, \bibinfo {author} {\bibfnamefont
  {J.}~\bibnamefont {L{\'e}onard}},\ and\ \bibinfo {author} {\bibfnamefont
  {M.}~\bibnamefont {Greiner}},\ }\bibfield  {title} {\bibinfo {title}
  {{Quantum critical behaviour at the many-body localization transition}},\
  }\href {https://doi.org/10.1038/s41586-019-1527-2} {\bibfield  {journal}
  {\bibinfo  {journal} {Nature}\ }\textbf {\bibinfo {volume} {573}},\ \bibinfo
  {pages} {385–389} (\bibinfo {year} {2019})}\BibitemShut {NoStop}%
\bibitem [{\citenamefont {Léonard}\ \emph {et~al.}(2023)\citenamefont
  {Léonard}, \citenamefont {Kim}, \citenamefont {Rispoli}, \citenamefont
  {Lukin}, \citenamefont {Schittko}, \citenamefont {Kwan}, \citenamefont
  {Demler}, \citenamefont {Sels},\ and\ \citenamefont {Greiner}}]{Leonard2023}%
  \BibitemOpen
  \bibfield  {author} {\bibinfo {author} {\bibfnamefont {J.}~\bibnamefont
  {Léonard}}, \bibinfo {author} {\bibfnamefont {S.}~\bibnamefont {Kim}},
  \bibinfo {author} {\bibfnamefont {M.}~\bibnamefont {Rispoli}}, \bibinfo
  {author} {\bibfnamefont {A.}~\bibnamefont {Lukin}}, \bibinfo {author}
  {\bibfnamefont {R.}~\bibnamefont {Schittko}}, \bibinfo {author}
  {\bibfnamefont {J.}~\bibnamefont {Kwan}}, \bibinfo {author} {\bibfnamefont
  {E.}~\bibnamefont {Demler}}, \bibinfo {author} {\bibfnamefont
  {D.}~\bibnamefont {Sels}},\ and\ \bibinfo {author} {\bibfnamefont
  {M.}~\bibnamefont {Greiner}},\ }\bibfield  {title} {\bibinfo {title}
  {{Probing the onset of quantum avalanches in a many-body localized system}},\
  }\href {https://doi.org/10.1038/s41567-022-01887-3} {\bibfield  {journal}
  {\bibinfo  {journal} {Nat. Phys.}\ }\textbf {\bibinfo {volume} {19}},\
  \bibinfo {pages} {481–485} (\bibinfo {year} {2023})}\BibitemShut {NoStop}%
\bibitem [{\citenamefont {Smith}\ \emph {et~al.}(2016)\citenamefont {Smith},
  \citenamefont {Lee}, \citenamefont {Richerme}, \citenamefont {Neyenhuis},
  \citenamefont {Hess}, \citenamefont {Hauke}, \citenamefont {Heyl},
  \citenamefont {Huse},\ and\ \citenamefont {Monroe}}]{Smith2016}%
  \BibitemOpen
  \bibfield  {author} {\bibinfo {author} {\bibfnamefont {J.}~\bibnamefont
  {Smith}}, \bibinfo {author} {\bibfnamefont {A.}~\bibnamefont {Lee}}, \bibinfo
  {author} {\bibfnamefont {P.}~\bibnamefont {Richerme}}, \bibinfo {author}
  {\bibfnamefont {B.}~\bibnamefont {Neyenhuis}}, \bibinfo {author}
  {\bibfnamefont {P.~W.}\ \bibnamefont {Hess}}, \bibinfo {author}
  {\bibfnamefont {P.}~\bibnamefont {Hauke}}, \bibinfo {author} {\bibfnamefont
  {M.}~\bibnamefont {Heyl}}, \bibinfo {author} {\bibfnamefont {D.~A.}\
  \bibnamefont {Huse}},\ and\ \bibinfo {author} {\bibfnamefont
  {C.}~\bibnamefont {Monroe}},\ }\bibfield  {title} {\bibinfo {title}
  {{Many-body localization in a quantum simulator with programmable random
  disorder}},\ }\href {https://doi.org/10.1038/nphys3783} {\bibfield  {journal}
  {\bibinfo  {journal} {Nat. Phys.}\ }\textbf {\bibinfo {volume} {12}},\
  \bibinfo {pages} {907–911} (\bibinfo {year} {2016})}\BibitemShut {NoStop}%
\bibitem [{\citenamefont {Choi}\ \emph {et~al.}(2017)\citenamefont {Choi},
  \citenamefont {Choi}, \citenamefont {Landig}, \citenamefont {Kucsko},
  \citenamefont {Zhou}, \citenamefont {Isoya}, \citenamefont {Jelezko},
  \citenamefont {Onoda}, \citenamefont {Sumiya}, \citenamefont {Khemani},
  \citenamefont {von Keyserlingk}, \citenamefont {Yao}, \citenamefont
  {Demler},\ and\ \citenamefont {Lukin}}]{Choi2017}%
  \BibitemOpen
  \bibfield  {author} {\bibinfo {author} {\bibfnamefont {S.}~\bibnamefont
  {Choi}}, \bibinfo {author} {\bibfnamefont {J.}~\bibnamefont {Choi}}, \bibinfo
  {author} {\bibfnamefont {R.}~\bibnamefont {Landig}}, \bibinfo {author}
  {\bibfnamefont {G.}~\bibnamefont {Kucsko}}, \bibinfo {author} {\bibfnamefont
  {H.}~\bibnamefont {Zhou}}, \bibinfo {author} {\bibfnamefont {J.}~\bibnamefont
  {Isoya}}, \bibinfo {author} {\bibfnamefont {F.}~\bibnamefont {Jelezko}},
  \bibinfo {author} {\bibfnamefont {S.}~\bibnamefont {Onoda}}, \bibinfo
  {author} {\bibfnamefont {H.}~\bibnamefont {Sumiya}}, \bibinfo {author}
  {\bibfnamefont {V.}~\bibnamefont {Khemani}}, \bibinfo {author} {\bibfnamefont
  {C.}~\bibnamefont {von Keyserlingk}}, \bibinfo {author} {\bibfnamefont
  {N.~Y.}\ \bibnamefont {Yao}}, \bibinfo {author} {\bibfnamefont
  {E.}~\bibnamefont {Demler}},\ and\ \bibinfo {author} {\bibfnamefont {M.~D.}\
  \bibnamefont {Lukin}},\ }\bibfield  {title} {\bibinfo {title} {Observation of
  discrete time-crystalline order in a disordered dipolar many-body system},\
  }\href {https://doi.org/10.1038/nature21426} {\bibfield  {journal} {\bibinfo
  {journal} {Nature}\ }\textbf {\bibinfo {volume} {543}},\ \bibinfo {pages}
  {221–225} (\bibinfo {year} {2017})}\BibitemShut {NoStop}%
\bibitem [{\citenamefont {Roushan}\ \emph {et~al.}(2017)\citenamefont
  {Roushan}, \citenamefont {Neill}, \citenamefont {Tangpanitanon},
  \citenamefont {Bastidas}, \citenamefont {Megrant}, \citenamefont {Barends},
  \citenamefont {Chen}, \citenamefont {Chen}, \citenamefont {Chiaro},
  \citenamefont {Dunsworth}, \citenamefont {Fowler}, \citenamefont {Foxen},
  \citenamefont {Giustina}, \citenamefont {Jeffrey}, \citenamefont {Kelly},
  \citenamefont {Lucero}, \citenamefont {Mutus}, \citenamefont {Neeley},
  \citenamefont {Quintana}, \citenamefont {D.~Sank}, \citenamefont {Neven},
  \citenamefont {Angelakis},\ and\ \citenamefont {Martinis}}]{Roushan2017}%
  \BibitemOpen
  \bibfield  {author} {\bibinfo {author} {\bibfnamefont {P.}~\bibnamefont
  {Roushan}}, \bibinfo {author} {\bibfnamefont {C.}~\bibnamefont {Neill}},
  \bibinfo {author} {\bibfnamefont {J.}~\bibnamefont {Tangpanitanon}}, \bibinfo
  {author} {\bibfnamefont {V.~M.}\ \bibnamefont {Bastidas}}, \bibinfo {author}
  {\bibfnamefont {A.}~\bibnamefont {Megrant}}, \bibinfo {author} {\bibfnamefont
  {R.}~\bibnamefont {Barends}}, \bibinfo {author} {\bibfnamefont
  {Y.}~\bibnamefont {Chen}}, \bibinfo {author} {\bibfnamefont {Z.}~\bibnamefont
  {Chen}}, \bibinfo {author} {\bibfnamefont {B.}~\bibnamefont {Chiaro}},
  \bibinfo {author} {\bibfnamefont {A.}~\bibnamefont {Dunsworth}}, \bibinfo
  {author} {\bibfnamefont {A.}~\bibnamefont {Fowler}}, \bibinfo {author}
  {\bibfnamefont {B.}~\bibnamefont {Foxen}}, \bibinfo {author} {\bibfnamefont
  {M.}~\bibnamefont {Giustina}}, \bibinfo {author} {\bibfnamefont
  {E.}~\bibnamefont {Jeffrey}}, \bibinfo {author} {\bibfnamefont
  {J.}~\bibnamefont {Kelly}}, \bibinfo {author} {\bibfnamefont
  {E.}~\bibnamefont {Lucero}}, \bibinfo {author} {\bibfnamefont
  {J.}~\bibnamefont {Mutus}}, \bibinfo {author} {\bibfnamefont
  {M.}~\bibnamefont {Neeley}}, \bibinfo {author} {\bibfnamefont
  {C.}~\bibnamefont {Quintana}}, \bibinfo {author} {\bibfnamefont {J.~W.
  T.~W.}\ \bibnamefont {D.~Sank}, \bibfnamefont {A.~Vainsencher}}, \bibinfo
  {author} {\bibfnamefont {H.}~\bibnamefont {Neven}}, \bibinfo {author}
  {\bibfnamefont {D.~G.}\ \bibnamefont {Angelakis}},\ and\ \bibinfo {author}
  {\bibfnamefont {J.}~\bibnamefont {Martinis}},\ }\bibfield  {title} {\bibinfo
  {title} {{Spectroscopic signatures of localization with interacting photons
  in superconducting qubits}},\ }\href
  {https://doi.org/10.1126/science.aao1401} {\bibfield  {journal} {\bibinfo
  {journal} {Science}\ }\textbf {\bibinfo {volume} {358}},\ \bibinfo {pages}
  {1175–1179} (\bibinfo {year} {2017})}\BibitemShut {NoStop}%
\bibitem [{\citenamefont {Xu}\ \emph {et~al.}(2018)\citenamefont {Xu},
  \citenamefont {Chen}, \citenamefont {Zeng}, \citenamefont {Zhang},
  \citenamefont {Song}, \citenamefont {Liu}, \citenamefont {Guo}, \citenamefont
  {Zhang}, \citenamefont {Xu}, \citenamefont {Deng}, \citenamefont {Huang},
  \citenamefont {Wang}, \citenamefont {Zhu}, \citenamefont {Zheng},\ and\
  \citenamefont {Fan}}]{Xu2018}%
  \BibitemOpen
  \bibfield  {author} {\bibinfo {author} {\bibfnamefont {K.}~\bibnamefont
  {Xu}}, \bibinfo {author} {\bibfnamefont {J.-J.}\ \bibnamefont {Chen}},
  \bibinfo {author} {\bibfnamefont {Y.}~\bibnamefont {Zeng}}, \bibinfo {author}
  {\bibfnamefont {Y.-R.}\ \bibnamefont {Zhang}}, \bibinfo {author}
  {\bibfnamefont {C.}~\bibnamefont {Song}}, \bibinfo {author} {\bibfnamefont
  {W.}~\bibnamefont {Liu}}, \bibinfo {author} {\bibfnamefont {Q.}~\bibnamefont
  {Guo}}, \bibinfo {author} {\bibfnamefont {P.}~\bibnamefont {Zhang}}, \bibinfo
  {author} {\bibfnamefont {D.}~\bibnamefont {Xu}}, \bibinfo {author}
  {\bibfnamefont {H.}~\bibnamefont {Deng}}, \bibinfo {author} {\bibfnamefont
  {K.}~\bibnamefont {Huang}}, \bibinfo {author} {\bibfnamefont
  {H.}~\bibnamefont {Wang}}, \bibinfo {author} {\bibfnamefont {X.}~\bibnamefont
  {Zhu}}, \bibinfo {author} {\bibfnamefont {D.}~\bibnamefont {Zheng}},\ and\
  \bibinfo {author} {\bibfnamefont {H.}~\bibnamefont {Fan}},\ }\bibfield
  {title} {\bibinfo {title} {{Emulating Many-Body Localization with a
  Superconducting Quantum Processor}},\ }\href
  {https://doi.org/10.1103/PhysRevLett.120.050507} {\bibfield  {journal}
  {\bibinfo  {journal} {Phys. Rev. Lett.}\ }\textbf {\bibinfo {volume} {120}},\
  \bibinfo {pages} {050507} (\bibinfo {year} {2018})}\BibitemShut {NoStop}%
\bibitem [{\citenamefont {Guo}\ \emph {et~al.}(2019)\citenamefont {Guo},
  \citenamefont {Cheng}, \citenamefont {Sun}, \citenamefont {Song},
  \citenamefont {Li}, \citenamefont {Wang}, \citenamefont {Ren}, \citenamefont
  {Dong}, \citenamefont {Zheng}, \citenamefont {Zhang}, \citenamefont
  {Mondaini}, \citenamefont {Fan},\ and\ \citenamefont {Wang}}]{Guo2019}%
  \BibitemOpen
  \bibfield  {author} {\bibinfo {author} {\bibfnamefont {Q.}~\bibnamefont
  {Guo}}, \bibinfo {author} {\bibfnamefont {C.}~\bibnamefont {Cheng}}, \bibinfo
  {author} {\bibfnamefont {Z.-H.}\ \bibnamefont {Sun}}, \bibinfo {author}
  {\bibfnamefont {Z.}~\bibnamefont {Song}}, \bibinfo {author} {\bibfnamefont
  {H.}~\bibnamefont {Li}}, \bibinfo {author} {\bibfnamefont {Z.}~\bibnamefont
  {Wang}}, \bibinfo {author} {\bibfnamefont {W.}~\bibnamefont {Ren}}, \bibinfo
  {author} {\bibfnamefont {H.}~\bibnamefont {Dong}}, \bibinfo {author}
  {\bibfnamefont {D.}~\bibnamefont {Zheng}}, \bibinfo {author} {\bibfnamefont
  {Y.-R.}\ \bibnamefont {Zhang}}, \bibinfo {author} {\bibfnamefont
  {R.}~\bibnamefont {Mondaini}}, \bibinfo {author} {\bibfnamefont
  {H.}~\bibnamefont {Fan}},\ and\ \bibinfo {author} {\bibfnamefont
  {H.}~\bibnamefont {Wang}},\ }\bibfield  {title} {\bibinfo {title}
  {{Observation of energy-resolved many-body localization}},\ }\href
  {https://doi.org/10.1038/s41567-020-1035-1} {\bibfield  {journal} {\bibinfo
  {journal} {Nat. Phys.}\ }\textbf {\bibinfo {volume} {17}},\ \bibinfo {pages}
  {234–239} (\bibinfo {year} {2019})}\BibitemShut {NoStop}%
\bibitem [{\citenamefont {Yao}\ \emph {et~al.}(2023)\citenamefont {Yao},
  \citenamefont {Xiang}, \citenamefont {Guo}, \citenamefont {Bao},
  \citenamefont {Yang}, \citenamefont {Song}, \citenamefont {Shi},
  \citenamefont {Zhu}, \citenamefont {Jin}, \citenamefont {Chen}, \citenamefont
  {Xu}, \citenamefont {Zhu}, \citenamefont {Shen}, \citenamefont {Wang},
  \citenamefont {Zhang}, \citenamefont {Wu}, \citenamefont {Zou}, \citenamefont
  {Zhang}, \citenamefont {Li}, \citenamefont {Wang}, \citenamefont {Song},
  \citenamefont {Cheng}, \citenamefont {Mondaini}, \citenamefont {Wang},
  \citenamefont {You}, \citenamefont {Zhu}, \citenamefont {Ying},\ and\
  \citenamefont {Guo}}]{Yao2023}%
  \BibitemOpen
  \bibfield  {author} {\bibinfo {author} {\bibfnamefont {Y.}~\bibnamefont
  {Yao}}, \bibinfo {author} {\bibfnamefont {L.}~\bibnamefont {Xiang}}, \bibinfo
  {author} {\bibfnamefont {Z.}~\bibnamefont {Guo}}, \bibinfo {author}
  {\bibfnamefont {Z.}~\bibnamefont {Bao}}, \bibinfo {author} {\bibfnamefont
  {Y.-F.}\ \bibnamefont {Yang}}, \bibinfo {author} {\bibfnamefont
  {Z.}~\bibnamefont {Song}}, \bibinfo {author} {\bibfnamefont {H.}~\bibnamefont
  {Shi}}, \bibinfo {author} {\bibfnamefont {X.}~\bibnamefont {Zhu}}, \bibinfo
  {author} {\bibfnamefont {F.}~\bibnamefont {Jin}}, \bibinfo {author}
  {\bibfnamefont {J.}~\bibnamefont {Chen}}, \bibinfo {author} {\bibfnamefont
  {S.}~\bibnamefont {Xu}}, \bibinfo {author} {\bibfnamefont {Z.}~\bibnamefont
  {Zhu}}, \bibinfo {author} {\bibfnamefont {F.}~\bibnamefont {Shen}}, \bibinfo
  {author} {\bibfnamefont {N.}~\bibnamefont {Wang}}, \bibinfo {author}
  {\bibfnamefont {C.}~\bibnamefont {Zhang}}, \bibinfo {author} {\bibfnamefont
  {Y.}~\bibnamefont {Wu}}, \bibinfo {author} {\bibfnamefont {Y.}~\bibnamefont
  {Zou}}, \bibinfo {author} {\bibfnamefont {P.}~\bibnamefont {Zhang}}, \bibinfo
  {author} {\bibfnamefont {H.}~\bibnamefont {Li}}, \bibinfo {author}
  {\bibfnamefont {Z.}~\bibnamefont {Wang}}, \bibinfo {author} {\bibfnamefont
  {C.}~\bibnamefont {Song}}, \bibinfo {author} {\bibfnamefont {C.}~\bibnamefont
  {Cheng}}, \bibinfo {author} {\bibfnamefont {R.}~\bibnamefont {Mondaini}},
  \bibinfo {author} {\bibfnamefont {H.}~\bibnamefont {Wang}}, \bibinfo {author}
  {\bibfnamefont {J.~Q.}\ \bibnamefont {You}}, \bibinfo {author} {\bibfnamefont
  {S.-Y.}\ \bibnamefont {Zhu}}, \bibinfo {author} {\bibfnamefont
  {L.}~\bibnamefont {Ying}},\ and\ \bibinfo {author} {\bibfnamefont
  {Q.}~\bibnamefont {Guo}},\ }\bibfield  {title} {\bibinfo {title}
  {{Observation of many-body Fock space dynamics in two dimensions}},\ }\href
  {https://doi.org/10.1038/s41567-023-02133-0} {\bibfield  {journal} {\bibinfo
  {journal} {Nat. Phys.}\ }\textbf {\bibinfo {volume} {19}},\ \bibinfo {pages}
  {1459–1465} (\bibinfo {year} {2023})}\BibitemShut {NoStop}%
\bibitem [{\citenamefont {De~Roeck}\ and\ \citenamefont
  {Huveneers}(2017)}]{DeRoeck2017}%
  \BibitemOpen
  \bibfield  {author} {\bibinfo {author} {\bibfnamefont {W.}~\bibnamefont
  {De~Roeck}}\ and\ \bibinfo {author} {\bibfnamefont {F.}~\bibnamefont
  {Huveneers}},\ }\bibfield  {title} {\bibinfo {title} {{Stability and
  instability towards delocalization in many-body localization systems}},\
  }\href {https://doi.org/10.1103/PhysRevB.95.155129} {\bibfield  {journal}
  {\bibinfo  {journal} {Phys. Rev. B}\ }\textbf {\bibinfo {volume} {95}},\
  \bibinfo {pages} {155129} (\bibinfo {year} {2017})}\BibitemShut {NoStop}%
\bibitem [{\citenamefont {Gopalakrishnan}\ and\ \citenamefont
  {Huse}(2019)}]{Gopalakrishnan2019}%
  \BibitemOpen
  \bibfield  {author} {\bibinfo {author} {\bibfnamefont {S.}~\bibnamefont
  {Gopalakrishnan}}\ and\ \bibinfo {author} {\bibfnamefont {D.~A.}\
  \bibnamefont {Huse}},\ }\bibfield  {title} {\bibinfo {title} {{Instability of
  many-body localized systems as a phase transition in a nonstandard
  thermodynamic limit}},\ }\href {https://doi.org/10.1103/PhysRevB.99.134305}
  {\bibfield  {journal} {\bibinfo  {journal} {Phys. Rev. B}\ }\textbf {\bibinfo
  {volume} {99}},\ \bibinfo {pages} {134305} (\bibinfo {year}
  {2019})}\BibitemShut {NoStop}%
\bibitem [{\citenamefont {{\v{S}}untajs}\ \emph {et~al.}(2020)\citenamefont
  {{\v{S}}untajs}, \citenamefont {Bon\v{c}a}, \citenamefont {Prosen},\ and\
  \citenamefont {Vidmar}}]{Suntajs2020}%
  \BibitemOpen
  \bibfield  {author} {\bibinfo {author} {\bibfnamefont {J.}~\bibnamefont
  {{\v{S}}untajs}}, \bibinfo {author} {\bibfnamefont {J.}~\bibnamefont
  {Bon\v{c}a}}, \bibinfo {author} {\bibfnamefont {T.}~\bibnamefont {Prosen}},\
  and\ \bibinfo {author} {\bibfnamefont {L.}~\bibnamefont {Vidmar}},\
  }\bibfield  {title} {\bibinfo {title} {{Quantum chaos challenges many-body
  localization}},\ }\href {https://doi.org/10.1103/PhysRevE.102.062144}
  {\bibfield  {journal} {\bibinfo  {journal} {Phys. Rev. E}\ }\textbf {\bibinfo
  {volume} {102}},\ \bibinfo {pages} {062144} (\bibinfo {year}
  {2020})}\BibitemShut {NoStop}%
\bibitem [{\citenamefont {Sels}\ and\ \citenamefont
  {Polkovnikov}(2021)}]{Sels2021}%
  \BibitemOpen
  \bibfield  {author} {\bibinfo {author} {\bibfnamefont {D.}~\bibnamefont
  {Sels}}\ and\ \bibinfo {author} {\bibfnamefont {A.}~\bibnamefont
  {Polkovnikov}},\ }\bibfield  {title} {\bibinfo {title} {{Dynamical
  obstruction to localization in a disordered spin chain}},\ }\href
  {https://doi.org/10.1103/PhysRevE.104.054105} {\bibfield  {journal} {\bibinfo
   {journal} {Phys. Rev. E}\ }\textbf {\bibinfo {volume} {104}},\ \bibinfo
  {pages} {054105} (\bibinfo {year} {2021})}\BibitemShut {NoStop}%
\bibitem [{\citenamefont {Peacock}\ and\ \citenamefont
  {Sels}(2023)}]{Peacock2023}%
  \BibitemOpen
  \bibfield  {author} {\bibinfo {author} {\bibfnamefont {J.~C.}\ \bibnamefont
  {Peacock}}\ and\ \bibinfo {author} {\bibfnamefont {D.}~\bibnamefont {Sels}},\
  }\bibfield  {title} {\bibinfo {title} {{Many-body delocalization from
  embedded thermal inclusion}},\ }\href
  {https://doi.org/10.1103/PhysRevB.108.L020201} {\bibfield  {journal}
  {\bibinfo  {journal} {Phys. Rev. B}\ }\textbf {\bibinfo {volume} {108}},\
  \bibinfo {pages} {L020201} (\bibinfo {year} {2023})}\BibitemShut {NoStop}%
\bibitem [{\citenamefont {Prelovšek}\ \emph {et~al.}(2021)\citenamefont
  {Prelovšek}, \citenamefont {Mierzejewski}, \citenamefont {Krsnik},\ and\
  \citenamefont {Barišić}}]{Prelovsek2021}%
  \BibitemOpen
  \bibfield  {author} {\bibinfo {author} {\bibfnamefont {P.}~\bibnamefont
  {Prelovšek}}, \bibinfo {author} {\bibfnamefont {M.}~\bibnamefont
  {Mierzejewski}}, \bibinfo {author} {\bibfnamefont {J.}~\bibnamefont
  {Krsnik}},\ and\ \bibinfo {author} {\bibfnamefont {O.~S.}\ \bibnamefont
  {Barišić}},\ }\bibfield  {title} {\bibinfo {title} {{Many-body localization
  as a percolation phenomenon}},\ }\href
  {https://doi.org/10.1103/PhysRevB.103.045139} {\bibfield  {journal} {\bibinfo
   {journal} {Phys. Rev. B}\ }\textbf {\bibinfo {volume} {103}},\ \bibinfo
  {pages} {045139} (\bibinfo {year} {2021})}\BibitemShut {NoStop}%
\bibitem [{\citenamefont {Rubio-Abadal}\ \emph {et~al.}(2019)\citenamefont
  {Rubio-Abadal}, \citenamefont {Choi}, \citenamefont {Zeiher}, \citenamefont
  {Hollerith}, \citenamefont {Rui}, \citenamefont {Bloch},\ and\ \citenamefont
  {Gross}}]{Rubio2019}%
  \BibitemOpen
  \bibfield  {author} {\bibinfo {author} {\bibfnamefont {A.}~\bibnamefont
  {Rubio-Abadal}}, \bibinfo {author} {\bibfnamefont {J.-y.}\ \bibnamefont
  {Choi}}, \bibinfo {author} {\bibfnamefont {J.}~\bibnamefont {Zeiher}},
  \bibinfo {author} {\bibfnamefont {S.}~\bibnamefont {Hollerith}}, \bibinfo
  {author} {\bibfnamefont {J.}~\bibnamefont {Rui}}, \bibinfo {author}
  {\bibfnamefont {I.}~\bibnamefont {Bloch}},\ and\ \bibinfo {author}
  {\bibfnamefont {C.}~\bibnamefont {Gross}},\ }\bibfield  {title} {\bibinfo
  {title} {{Many-Body Delocalization in the Presence of a Quantum Bath}},\
  }\href {https://doi.org/10.1103/PhysRevX.9.041014} {\bibfield  {journal}
  {\bibinfo  {journal} {Phys. Rev. X}\ }\textbf {\bibinfo {volume} {9}},\
  \bibinfo {pages} {041014} (\bibinfo {year} {2019})}\BibitemShut {NoStop}%
\bibitem [{\citenamefont {Wahl}\ \emph {et~al.}(2019)\citenamefont {Wahl},
  \citenamefont {Pal},\ and\ \citenamefont {Simon}}]{Wahl2019}%
  \BibitemOpen
  \bibfield  {author} {\bibinfo {author} {\bibfnamefont {T.~B.}\ \bibnamefont
  {Wahl}}, \bibinfo {author} {\bibfnamefont {A.}~\bibnamefont {Pal}},\ and\
  \bibinfo {author} {\bibfnamefont {S.~H.}\ \bibnamefont {Simon}},\ }\bibfield
  {title} {\bibinfo {title} {{Signatures of the many-body localized regime in
  two dimensions}},\ }\href {https://doi.org/10.1038/s41567-018-0339-x}
  {\bibfield  {journal} {\bibinfo  {journal} {Nat. Phys.}\ }\textbf {\bibinfo
  {volume} {15}},\ \bibinfo {pages} {164–169} (\bibinfo {year}
  {2019})}\BibitemShut {NoStop}%
\bibitem [{\citenamefont {Chertkov}\ \emph {et~al.}(2021)\citenamefont
  {Chertkov}, \citenamefont {Villalonga},\ and\ \citenamefont
  {Clark}}]{Chertkov2021}%
  \BibitemOpen
  \bibfield  {author} {\bibinfo {author} {\bibfnamefont {E.}~\bibnamefont
  {Chertkov}}, \bibinfo {author} {\bibfnamefont {B.}~\bibnamefont
  {Villalonga}},\ and\ \bibinfo {author} {\bibfnamefont {B.~K.}\ \bibnamefont
  {Clark}},\ }\bibfield  {title} {\bibinfo {title} {{Numerical Evidence for
  Many-Body Localization in Two and Three Dimensions}},\ }\href
  {https://doi.org/10.1103/PhysRevLett.126.180602} {\bibfield  {journal}
  {\bibinfo  {journal} {Phys. Rev. Lett.}\ }\textbf {\bibinfo {volume} {126}},\
  \bibinfo {pages} {180602} (\bibinfo {year} {2021})}\BibitemShut {NoStop}%
\bibitem [{\citenamefont {Doggen}\ \emph {et~al.}(2020)\citenamefont {Doggen},
  \citenamefont {Gornyi}, \citenamefont {Mirlin},\ and\ \citenamefont
  {Polyakov}}]{Doggen2020}%
  \BibitemOpen
  \bibfield  {author} {\bibinfo {author} {\bibfnamefont {E.~V.~H.}\
  \bibnamefont {Doggen}}, \bibinfo {author} {\bibfnamefont {I.~V.}\
  \bibnamefont {Gornyi}}, \bibinfo {author} {\bibfnamefont {A.~D.}\
  \bibnamefont {Mirlin}},\ and\ \bibinfo {author} {\bibfnamefont {D.~G.}\
  \bibnamefont {Polyakov}},\ }\bibfield  {title} {\bibinfo {title} {{Slow
  Many-Body Delocalization beyond One Dimension}},\ }\href
  {https://doi.org/10.1103/PhysRevLett.125.155701} {\bibfield  {journal}
  {\bibinfo  {journal} {Phys. Rev. Lett.}\ }\textbf {\bibinfo {volume} {125}},\
  \bibinfo {pages} {155701} (\bibinfo {year} {2020})}\BibitemShut {NoStop}%
\bibitem [{\citenamefont {Li}\ \emph {et~al.}(2024)\citenamefont {Li},
  \citenamefont {Chan},\ and\ \citenamefont {Wahl}}]{Li2024}%
  \BibitemOpen
  \bibfield  {author} {\bibinfo {author} {\bibfnamefont {J.}~\bibnamefont
  {Li}}, \bibinfo {author} {\bibfnamefont {A.}~\bibnamefont {Chan}},\ and\
  \bibinfo {author} {\bibfnamefont {T.~B.}\ \bibnamefont {Wahl}},\ }\bibfield
  {title} {\bibinfo {title} {{Quantum circuits reproduce the experimental
  two-dimensional many-body localization transition point}},\ }\href
  {https://doi.org/10.1103/PhysRevB.109.L140202} {\bibfield  {journal}
  {\bibinfo  {journal} {Phys. Rev. B}\ }\textbf {\bibinfo {volume} {109}},\
  \bibinfo {pages} {L140202} (\bibinfo {year} {2024})}\BibitemShut {NoStop}%
\bibitem [{\citenamefont {Li}\ \emph {et~al.}(2025{\natexlab{a}})\citenamefont
  {Li}, \citenamefont {Chan},\ and\ \citenamefont {Wahl}}]{Li2025}%
  \BibitemOpen
  \bibfield  {author} {\bibinfo {author} {\bibfnamefont {J.}~\bibnamefont
  {Li}}, \bibinfo {author} {\bibfnamefont {A.}~\bibnamefont {Chan}},\ and\
  \bibinfo {author} {\bibfnamefont {T.~B.}\ \bibnamefont {Wahl}},\ }\bibfield
  {title} {\bibinfo {title} {{Two-dimensional many-body localized systems
  coupled to a heat bath}},\ }\href {https://doi.org/10.1103/7wld-nzbn}
  {\bibfield  {journal} {\bibinfo  {journal} {Phys. Rev. B}\ }\textbf {\bibinfo
  {volume} {111}},\ \bibinfo {pages} {224211} (\bibinfo {year}
  {2025}{\natexlab{a}})}\BibitemShut {NoStop}%
\bibitem [{\citenamefont {Agrawal}\ \emph {et~al.}(2022)\citenamefont
  {Agrawal}, \citenamefont {Vasseur},\ and\ \citenamefont
  {Gopalakrishnan}}]{Agrawal2022}%
  \BibitemOpen
  \bibfield  {author} {\bibinfo {author} {\bibfnamefont {U.}~\bibnamefont
  {Agrawal}}, \bibinfo {author} {\bibfnamefont {R.}~\bibnamefont {Vasseur}},\
  and\ \bibinfo {author} {\bibfnamefont {S.}~\bibnamefont {Gopalakrishnan}},\
  }\bibfield  {title} {\bibinfo {title} {{Quasiperiodic many-body localization
  transition in dimension $d>1$}},\ }\href
  {https://doi.org/10.1103/PhysRevB.106.094206} {\bibfield  {journal} {\bibinfo
   {journal} {Phys. Rev. B}\ }\textbf {\bibinfo {volume} {106}},\ \bibinfo
  {pages} {094206} (\bibinfo {year} {2022})}\BibitemShut {NoStop}%
\bibitem [{\citenamefont {{\v{S}}trkalj}\ \emph {et~al.}(2022)\citenamefont
  {{\v{S}}trkalj}, \citenamefont {Doggen},\ and\ \citenamefont
  {Castelnovo}}]{Strkalj2022}%
  \BibitemOpen
  \bibfield  {author} {\bibinfo {author} {\bibfnamefont {A.}~\bibnamefont
  {{\v{S}}trkalj}}, \bibinfo {author} {\bibfnamefont {E.~V.~H.}\ \bibnamefont
  {Doggen}},\ and\ \bibinfo {author} {\bibfnamefont {C.}~\bibnamefont
  {Castelnovo}},\ }\bibfield  {title} {\bibinfo {title} {{Coexistence of
  localization and transport in many-body two-dimensional Aubry-Andr\'e
  models}},\ }\href {https://doi.org/10.1103/PhysRevB.106.184209} {\bibfield
  {journal} {\bibinfo  {journal} {Phys. Rev. B}\ }\textbf {\bibinfo {volume}
  {106}},\ \bibinfo {pages} {184209} (\bibinfo {year} {2022})}\BibitemShut
  {NoStop}%
\bibitem [{\citenamefont {Crowley}\ and\ \citenamefont
  {Chandran}(2022)}]{Crowley2022}%
  \BibitemOpen
  \bibfield  {author} {\bibinfo {author} {\bibfnamefont {P.~J.~D.}\
  \bibnamefont {Crowley}}\ and\ \bibinfo {author} {\bibfnamefont
  {A.}~\bibnamefont {Chandran}},\ }\bibfield  {title} {\bibinfo {title}
  {{Mean-field theory of failed thermalizing avalanches}},\ }\href
  {https://doi.org/10.1103/PhysRevB.106.184208} {\bibfield  {journal} {\bibinfo
   {journal} {Phys. Rev. B}\ }\textbf {\bibinfo {volume} {106}},\ \bibinfo
  {pages} {184208} (\bibinfo {year} {2022})}\BibitemShut {NoStop}%
\bibitem [{\citenamefont {Potirniche}\ \emph {et~al.}(2019)\citenamefont
  {Potirniche}, \citenamefont {Banerjee},\ and\ \citenamefont
  {Altman}}]{Potirniche2019}%
  \BibitemOpen
  \bibfield  {author} {\bibinfo {author} {\bibfnamefont {I.-D.}\ \bibnamefont
  {Potirniche}}, \bibinfo {author} {\bibfnamefont {S.}~\bibnamefont
  {Banerjee}},\ and\ \bibinfo {author} {\bibfnamefont {E.}~\bibnamefont
  {Altman}},\ }\bibfield  {title} {\bibinfo {title} {{Exploration of the
  stability of many-body localization in $d>1$}},\ }\href
  {https://doi.org/10.1103/PhysRevB.99.205149} {\bibfield  {journal} {\bibinfo
  {journal} {Phys. Rev. B}\ }\textbf {\bibinfo {volume} {99}},\ \bibinfo
  {pages} {205149} (\bibinfo {year} {2019})}\BibitemShut {NoStop}%
\bibitem [{\citenamefont {Haegeman}\ \emph {et~al.}(2016)\citenamefont
  {Haegeman}, \citenamefont {Lubich}, \citenamefont {Oseledets}, \citenamefont
  {Vandereycken},\ and\ \citenamefont {Verstraete}}]{tdvp}%
  \BibitemOpen
  \bibfield  {author} {\bibinfo {author} {\bibfnamefont {J.}~\bibnamefont
  {Haegeman}}, \bibinfo {author} {\bibfnamefont {C.}~\bibnamefont {Lubich}},
  \bibinfo {author} {\bibfnamefont {I.}~\bibnamefont {Oseledets}}, \bibinfo
  {author} {\bibfnamefont {B.}~\bibnamefont {Vandereycken}},\ and\ \bibinfo
  {author} {\bibfnamefont {F.}~\bibnamefont {Verstraete}},\ }\bibfield  {title}
  {\bibinfo {title} {Unifying time evolution and optimization with matrix
  product states},\ }\href {https://doi.org/10.1103/PhysRevB.94.165116}
  {\bibfield  {journal} {\bibinfo  {journal} {Phys. Rev. B}\ }\textbf {\bibinfo
  {volume} {94}},\ \bibinfo {pages} {165116} (\bibinfo {year}
  {2016})}\BibitemShut {NoStop}%
\bibitem [{\citenamefont {Billy}\ \emph {et~al.}(2008)\citenamefont {Billy},
  \citenamefont {Josse}, \citenamefont {Zuo}, \citenamefont {Bernard},
  \citenamefont {Hambrecht}, \citenamefont {Lugan}, \citenamefont {Clément},
  \citenamefont {Sanchez-Palencia}, \citenamefont {Bouyer},\ and\ \citenamefont
  {Aspect}}]{Billy2008}%
  \BibitemOpen
  \bibfield  {author} {\bibinfo {author} {\bibfnamefont {J.}~\bibnamefont
  {Billy}}, \bibinfo {author} {\bibfnamefont {V.}~\bibnamefont {Josse}},
  \bibinfo {author} {\bibfnamefont {Z.}~\bibnamefont {Zuo}}, \bibinfo {author}
  {\bibfnamefont {A.}~\bibnamefont {Bernard}}, \bibinfo {author} {\bibfnamefont
  {B.}~\bibnamefont {Hambrecht}}, \bibinfo {author} {\bibfnamefont
  {P.}~\bibnamefont {Lugan}}, \bibinfo {author} {\bibfnamefont
  {D.}~\bibnamefont {Clément}}, \bibinfo {author} {\bibfnamefont
  {L.}~\bibnamefont {Sanchez-Palencia}}, \bibinfo {author} {\bibfnamefont
  {P.}~\bibnamefont {Bouyer}},\ and\ \bibinfo {author} {\bibfnamefont
  {A.}~\bibnamefont {Aspect}},\ }\bibfield  {title} {\bibinfo {title} {{Direct
  observation of Anderson localization of matter waves in a controlled
  disorder}},\ }\href {https://doi.org/10.1038/nature07000} {\bibfield
  {journal} {\bibinfo  {journal} {Nature}\ }\textbf {\bibinfo {volume} {453}},\
  \bibinfo {pages} {891–894} (\bibinfo {year} {2008})}\BibitemShut {NoStop}%
\bibitem [{\citenamefont {Roati}\ \emph {et~al.}(2008)\citenamefont {Roati},
  \citenamefont {D'Errico}, \citenamefont {Fallani}, \citenamefont {Fattori},
  \citenamefont {Fort}, \citenamefont {Zaccanti}, \citenamefont {Modugno},
  \citenamefont {Modugno},\ and\ \citenamefont {Inguscio}}]{Roati2008}%
  \BibitemOpen
  \bibfield  {author} {\bibinfo {author} {\bibfnamefont {G.}~\bibnamefont
  {Roati}}, \bibinfo {author} {\bibfnamefont {C.}~\bibnamefont {D'Errico}},
  \bibinfo {author} {\bibfnamefont {L.}~\bibnamefont {Fallani}}, \bibinfo
  {author} {\bibfnamefont {M.}~\bibnamefont {Fattori}}, \bibinfo {author}
  {\bibfnamefont {C.}~\bibnamefont {Fort}}, \bibinfo {author} {\bibfnamefont
  {M.}~\bibnamefont {Zaccanti}}, \bibinfo {author} {\bibfnamefont
  {G.}~\bibnamefont {Modugno}}, \bibinfo {author} {\bibfnamefont
  {M.}~\bibnamefont {Modugno}},\ and\ \bibinfo {author} {\bibfnamefont
  {M.}~\bibnamefont {Inguscio}},\ }\bibfield  {title} {\bibinfo {title}
  {{Anderson localization of a non-interacting Bose–Einstein condensate}},\
  }\href {https://doi.org/10.1038/nature07071} {\bibfield  {journal} {\bibinfo
  {journal} {Nature}\ }\textbf {\bibinfo {volume} {453}},\ \bibinfo {pages}
  {895–898} (\bibinfo {year} {2008})}\BibitemShut {NoStop}%
\bibitem [{\citenamefont {Kondov}\ \emph {et~al.}(2011)\citenamefont {Kondov},
  \citenamefont {McGehee}, \citenamefont {Zirbel},\ and\ \citenamefont
  {DeMarco}}]{Kondov2011}%
  \BibitemOpen
  \bibfield  {author} {\bibinfo {author} {\bibfnamefont {S.~S.}\ \bibnamefont
  {Kondov}}, \bibinfo {author} {\bibfnamefont {W.~R.}\ \bibnamefont {McGehee}},
  \bibinfo {author} {\bibfnamefont {J.~J.}\ \bibnamefont {Zirbel}},\ and\
  \bibinfo {author} {\bibfnamefont {B.}~\bibnamefont {DeMarco}},\ }\bibfield
  {title} {\bibinfo {title} {{Three-dimensional Anderson localization of
  ultracold matter}},\ }\href {https://doi.org/10.1126/science.1209019}
  {\bibfield  {journal} {\bibinfo  {journal} {Science}\ }\textbf {\bibinfo
  {volume} {334}},\ \bibinfo {pages} {66–68} (\bibinfo {year}
  {2011})}\BibitemShut {NoStop}%
\bibitem [{\citenamefont {Yu}\ \emph {et~al.}(2024)\citenamefont {Yu},
  \citenamefont {Bhave}, \citenamefont {Reeve}, \citenamefont {Song},\ and\
  \citenamefont {Schneider}}]{Yu2024}%
  \BibitemOpen
  \bibfield  {author} {\bibinfo {author} {\bibfnamefont {J.-C.}\ \bibnamefont
  {Yu}}, \bibinfo {author} {\bibfnamefont {S.}~\bibnamefont {Bhave}}, \bibinfo
  {author} {\bibfnamefont {L.}~\bibnamefont {Reeve}}, \bibinfo {author}
  {\bibfnamefont {B.}~\bibnamefont {Song}},\ and\ \bibinfo {author}
  {\bibfnamefont {U.}~\bibnamefont {Schneider}},\ }\bibfield  {title} {\bibinfo
  {title} {Observation of discrete time-crystalline order in a disordered
  dipolar many-body system},\ }\href {https://doi.org/10.1038/nature21426}
  {\bibfield  {journal} {\bibinfo  {journal} {Nature}\ }\textbf {\bibinfo
  {volume} {633}},\ \bibinfo {pages} {338–343} (\bibinfo {year}
  {2024})}\BibitemShut {NoStop}%
\bibitem [{\citenamefont {Fukuhara}\ \emph {et~al.}(2013)\citenamefont
  {Fukuhara}, \citenamefont {Kantian}, \citenamefont {Endres}, \citenamefont
  {Cheneau}, \citenamefont {Schauß}, \citenamefont {Hild}, \citenamefont
  {Bellem}, \citenamefont {Schollwöck}, \citenamefont {Giamarchi},
  \citenamefont {Gross}, \citenamefont {Bloch},\ and\ \citenamefont
  {Kuhr}}]{Fukuhara2013}%
  \BibitemOpen
  \bibfield  {author} {\bibinfo {author} {\bibfnamefont {T.}~\bibnamefont
  {Fukuhara}}, \bibinfo {author} {\bibfnamefont {A.}~\bibnamefont {Kantian}},
  \bibinfo {author} {\bibfnamefont {M.}~\bibnamefont {Endres}}, \bibinfo
  {author} {\bibfnamefont {M.}~\bibnamefont {Cheneau}}, \bibinfo {author}
  {\bibfnamefont {P.}~\bibnamefont {Schauß}}, \bibinfo {author} {\bibfnamefont
  {S.}~\bibnamefont {Hild}}, \bibinfo {author} {\bibfnamefont {D.}~\bibnamefont
  {Bellem}}, \bibinfo {author} {\bibfnamefont {U.}~\bibnamefont {Schollwöck}},
  \bibinfo {author} {\bibfnamefont {T.}~\bibnamefont {Giamarchi}}, \bibinfo
  {author} {\bibfnamefont {C.}~\bibnamefont {Gross}}, \bibinfo {author}
  {\bibfnamefont {I.}~\bibnamefont {Bloch}},\ and\ \bibinfo {author}
  {\bibfnamefont {S.}~\bibnamefont {Kuhr}},\ }\bibfield  {title} {\bibinfo
  {title} {{Quantum dynamics of a mobile spin impurity}},\ }\href
  {https://doi.org/10.1038/nphys2561} {\bibfield  {journal} {\bibinfo
  {journal} {Nat. Phys.}\ }\textbf {\bibinfo {volume} {9}},\ \bibinfo {pages}
  {235–241} (\bibinfo {year} {2013})}\BibitemShut {NoStop}%
\bibitem [{\citenamefont {Mazurenko}\ \emph {et~al.}(2017)\citenamefont
  {Mazurenko}, \citenamefont {Chiu}, \citenamefont {Ji}, \citenamefont
  {Parsons}, \citenamefont {Kan{\'a}sz-Nagy}, \citenamefont {Schmidt},
  \citenamefont {Grusdt}, \citenamefont {Demler}, \citenamefont {Greif},\ and\
  \citenamefont {Greiner}}]{Mazurenko2017}%
  \BibitemOpen
  \bibfield  {author} {\bibinfo {author} {\bibfnamefont {A.}~\bibnamefont
  {Mazurenko}}, \bibinfo {author} {\bibfnamefont {C.~S.}\ \bibnamefont {Chiu}},
  \bibinfo {author} {\bibfnamefont {G.}~\bibnamefont {Ji}}, \bibinfo {author}
  {\bibfnamefont {M.~F.}\ \bibnamefont {Parsons}}, \bibinfo {author}
  {\bibfnamefont {M.}~\bibnamefont {Kan{\'a}sz-Nagy}}, \bibinfo {author}
  {\bibfnamefont {R.}~\bibnamefont {Schmidt}}, \bibinfo {author} {\bibfnamefont
  {F.}~\bibnamefont {Grusdt}}, \bibinfo {author} {\bibfnamefont
  {E.}~\bibnamefont {Demler}}, \bibinfo {author} {\bibfnamefont
  {D.}~\bibnamefont {Greif}},\ and\ \bibinfo {author} {\bibfnamefont
  {M.}~\bibnamefont {Greiner}},\ }\bibfield  {title} {\bibinfo {title} {{A
  cold-atom Fermi–Hubbard antiferromagnet}},\ }\href
  {https://doi.org/10.1038/nature22362} {\bibfield  {journal} {\bibinfo
  {journal} {Nature}\ }\textbf {\bibinfo {volume} {545}},\ \bibinfo {pages}
  {462–466} (\bibinfo {year} {2017})}\BibitemShut {NoStop}%
\bibitem [{\citenamefont {Kwon}\ \emph {et~al.}(2022)\citenamefont {Kwon},
  \citenamefont {Kim}, \citenamefont {Hur}, \citenamefont {Huh},\ and\
  \citenamefont {Choi}}]{Kwon2022}%
  \BibitemOpen
  \bibfield  {author} {\bibinfo {author} {\bibfnamefont {K.}~\bibnamefont
  {Kwon}}, \bibinfo {author} {\bibfnamefont {K.}~\bibnamefont {Kim}}, \bibinfo
  {author} {\bibfnamefont {J.}~\bibnamefont {Hur}}, \bibinfo {author}
  {\bibfnamefont {S.}~\bibnamefont {Huh}},\ and\ \bibinfo {author}
  {\bibfnamefont {J.-y.}\ \bibnamefont {Choi}},\ }\bibfield  {title} {\bibinfo
  {title} {{Site-resolved imaging of a bosonic Mott insulator of $^{7}${Li}
  atoms}},\ }\href {https://doi.org/10.1103/PhysRevA.105.033323} {\bibfield
  {journal} {\bibinfo  {journal} {Phys. Rev. A}\ }\textbf {\bibinfo {volume}
  {105}},\ \bibinfo {pages} {033323} (\bibinfo {year} {2022})}\BibitemShut
  {NoStop}%
\bibitem [{\citenamefont {Morong}\ \emph {et~al.}(2021)\citenamefont {Morong},
  \citenamefont {Liu}, \citenamefont {Becker}, \citenamefont {Collins},
  \citenamefont {Feng}, \citenamefont {Kyprianidis}, \citenamefont {Pagano},
  \citenamefont {You}, \citenamefont {Gorshkov},\ and\ \citenamefont
  {Monroe}}]{Morong2021}%
  \BibitemOpen
  \bibfield  {author} {\bibinfo {author} {\bibfnamefont {W.}~\bibnamefont
  {Morong}}, \bibinfo {author} {\bibfnamefont {F.}~\bibnamefont {Liu}},
  \bibinfo {author} {\bibfnamefont {P.}~\bibnamefont {Becker}}, \bibinfo
  {author} {\bibfnamefont {K.~S.}\ \bibnamefont {Collins}}, \bibinfo {author}
  {\bibfnamefont {L.}~\bibnamefont {Feng}}, \bibinfo {author} {\bibfnamefont
  {A.}~\bibnamefont {Kyprianidis}}, \bibinfo {author} {\bibfnamefont
  {G.}~\bibnamefont {Pagano}}, \bibinfo {author} {\bibfnamefont
  {T.}~\bibnamefont {You}}, \bibinfo {author} {\bibfnamefont {A.~V.}\
  \bibnamefont {Gorshkov}},\ and\ \bibinfo {author} {\bibfnamefont
  {C.}~\bibnamefont {Monroe}},\ }\bibfield  {title} {\bibinfo {title}
  {{Observation of Stark many-body localization without disorder}},\ }\href
  {https://doi.org/10.1038/s41586-021-03988-0} {\bibfield  {journal} {\bibinfo
  {journal} {Nature}\ }\textbf {\bibinfo {volume} {599}},\ \bibinfo {pages}
  {393–398} (\bibinfo {year} {2021})}\BibitemShut {NoStop}%
\bibitem [{\citenamefont {Sierant}\ \emph {et~al.}(2025)\citenamefont
  {Sierant}, \citenamefont {Lewenstein}, \citenamefont {Scardicchio},
  \citenamefont {Vidmar},\ and\ \citenamefont {Zakrzewski}}]{Sierant2025}%
  \BibitemOpen
  \bibfield  {author} {\bibinfo {author} {\bibfnamefont {P.}~\bibnamefont
  {Sierant}}, \bibinfo {author} {\bibfnamefont {M.}~\bibnamefont {Lewenstein}},
  \bibinfo {author} {\bibfnamefont {A.}~\bibnamefont {Scardicchio}}, \bibinfo
  {author} {\bibfnamefont {L.}~\bibnamefont {Vidmar}},\ and\ \bibinfo {author}
  {\bibfnamefont {J.}~\bibnamefont {Zakrzewski}},\ }\bibfield  {title}
  {\bibinfo {title} {{Many-body localization in the age of classical
  computing}},\ }\href {https://doi.org/10.1088/1361-6633/ad9756} {\bibfield
  {journal} {\bibinfo  {journal} {Rep. Prog. Phys}\ }\textbf {\bibinfo {volume}
  {88}},\ \bibinfo {pages} {026502} (\bibinfo {year} {2025})}\BibitemShut
  {NoStop}%
\bibitem [{\citenamefont {Chandran}\ \emph {et~al.}(2016)\citenamefont
  {Chandran}, \citenamefont {Pal}, \citenamefont {Laumann},\ and\ \citenamefont
  {Scardicchio}}]{Chandran2016}%
  \BibitemOpen
  \bibfield  {author} {\bibinfo {author} {\bibfnamefont {A.}~\bibnamefont
  {Chandran}}, \bibinfo {author} {\bibfnamefont {A.}~\bibnamefont {Pal}},
  \bibinfo {author} {\bibfnamefont {C.~R.}\ \bibnamefont {Laumann}},\ and\
  \bibinfo {author} {\bibfnamefont {A.}~\bibnamefont {Scardicchio}},\
  }\bibfield  {title} {\bibinfo {title} {{Many-body localization beyond
  eigenstates in all dimensions}},\ }\href
  {https://doi.org/10.1103/PhysRevB.94.144203} {\bibfield  {journal} {\bibinfo
  {journal} {Phys. Rev. B}\ }\textbf {\bibinfo {volume} {94}},\ \bibinfo
  {pages} {144203} (\bibinfo {year} {2016})}\BibitemShut {NoStop}%
\bibitem [{\citenamefont {Khemani}\ \emph {et~al.}(2017)\citenamefont
  {Khemani}, \citenamefont {Sheng},\ and\ \citenamefont {Huse}}]{Khemani2017}%
  \BibitemOpen
  \bibfield  {author} {\bibinfo {author} {\bibfnamefont {V.}~\bibnamefont
  {Khemani}}, \bibinfo {author} {\bibfnamefont {D.~N.}\ \bibnamefont {Sheng}},\
  and\ \bibinfo {author} {\bibfnamefont {D.~A.}\ \bibnamefont {Huse}},\
  }\bibfield  {title} {\bibinfo {title} {{Two Universality Classes for the
  Many-Body Localization Transition}},\ }\href
  {https://doi.org/10.1103/PhysRevLett.119.075702} {\bibfield  {journal}
  {\bibinfo  {journal} {Phys. Rev. Lett.}\ }\textbf {\bibinfo {volume} {119}},\
  \bibinfo {pages} {075702} (\bibinfo {year} {2017})}\BibitemShut {NoStop}%
\bibitem [{\citenamefont {Agrawal}\ \emph {et~al.}(2020)\citenamefont
  {Agrawal}, \citenamefont {Gopalakrishnan},\ and\ \citenamefont
  {Vasseur}}]{Agrawal2020}%
  \BibitemOpen
  \bibfield  {author} {\bibinfo {author} {\bibfnamefont {U.}~\bibnamefont
  {Agrawal}}, \bibinfo {author} {\bibfnamefont {S.}~\bibnamefont
  {Gopalakrishnan}},\ and\ \bibinfo {author} {\bibfnamefont {R.}~\bibnamefont
  {Vasseur}},\ }\bibfield  {title} {\bibinfo {title} {{Universality and quantum
  criticality in quasiperiodic spin chains}},\ }\href
  {https://doi.org/10.1038/s41467-020-15760-5} {\bibfield  {journal} {\bibinfo
  {journal} {Nat. Commun.}\ }\textbf {\bibinfo {volume} {11}},\ \bibinfo
  {pages} {2225} (\bibinfo {year} {2020})}\BibitemShut {NoStop}%
\bibitem [{\citenamefont {Li}\ \emph {et~al.}(2025{\natexlab{b}})\citenamefont
  {Li}, \citenamefont {Sun}, \citenamefont {Shi}, \citenamefont {Bao},
  \citenamefont {Wang}, \citenamefont {Zhang}, \citenamefont {Liu},
  \citenamefont {Deng}, \citenamefont {Yu}, \citenamefont {Liu}, \citenamefont
  {Chen}, \citenamefont {Li}, \citenamefont {Li}, \citenamefont {Liu},
  \citenamefont {Zhou}, \citenamefont {Peng}, \citenamefont {Liu},
  \citenamefont {Wang}, \citenamefont {Xu}, \citenamefont {Zhao}, \citenamefont
  {He}, \citenamefont {Feng}, \citenamefont {Song}, \citenamefont {Fang},
  \citenamefont {Deng}, \citenamefont {Xu}, \citenamefont {Chen}, \citenamefont
  {zhou}, \citenamefont {Liang}, \citenamefont {Xiang}, \citenamefont {Xue},
  \citenamefont {Zheng}, \citenamefont {Huang}, \citenamefont {Wang},
  \citenamefont {Yu}, \citenamefont {Sierant}, \citenamefont {Xu},\ and\
  \citenamefont {Fan}}]{70qubit2dMBL}%
  \BibitemOpen
  \bibfield  {author} {\bibinfo {author} {\bibfnamefont {T.-M.}\ \bibnamefont
  {Li}}, \bibinfo {author} {\bibfnamefont {Z.-H.}\ \bibnamefont {Sun}},
  \bibinfo {author} {\bibfnamefont {Y.-H.}\ \bibnamefont {Shi}}, \bibinfo
  {author} {\bibfnamefont {Z.-T.}\ \bibnamefont {Bao}}, \bibinfo {author}
  {\bibfnamefont {Y.-Y.}\ \bibnamefont {Wang}}, \bibinfo {author}
  {\bibfnamefont {J.-C.}\ \bibnamefont {Zhang}}, \bibinfo {author}
  {\bibfnamefont {Y.}~\bibnamefont {Liu}}, \bibinfo {author} {\bibfnamefont
  {C.-L.}\ \bibnamefont {Deng}}, \bibinfo {author} {\bibfnamefont {Y.-H.}\
  \bibnamefont {Yu}}, \bibinfo {author} {\bibfnamefont {Z.-H.}\ \bibnamefont
  {Liu}}, \bibinfo {author} {\bibfnamefont {C.-T.}\ \bibnamefont {Chen}},
  \bibinfo {author} {\bibfnamefont {L.}~\bibnamefont {Li}}, \bibinfo {author}
  {\bibfnamefont {H.}~\bibnamefont {Li}}, \bibinfo {author} {\bibfnamefont
  {H.-T.}\ \bibnamefont {Liu}}, \bibinfo {author} {\bibfnamefont {S.-Y.}\
  \bibnamefont {Zhou}}, \bibinfo {author} {\bibfnamefont {Z.-Y.}\ \bibnamefont
  {Peng}}, \bibinfo {author} {\bibfnamefont {Y.-J.}\ \bibnamefont {Liu}},
  \bibinfo {author} {\bibfnamefont {Z.}~\bibnamefont {Wang}}, \bibinfo {author}
  {\bibfnamefont {Y.-S.}\ \bibnamefont {Xu}}, \bibinfo {author} {\bibfnamefont
  {K.}~\bibnamefont {Zhao}}, \bibinfo {author} {\bibfnamefont {Y.}~\bibnamefont
  {He}}, \bibinfo {author} {\bibfnamefont {D.}~\bibnamefont {Feng}}, \bibinfo
  {author} {\bibfnamefont {J.-C.}\ \bibnamefont {Song}}, \bibinfo {author}
  {\bibfnamefont {C.-P.}\ \bibnamefont {Fang}}, \bibinfo {author}
  {\bibfnamefont {J.}~\bibnamefont {Deng}}, \bibinfo {author} {\bibfnamefont
  {M.}~\bibnamefont {Xu}}, \bibinfo {author} {\bibfnamefont {Y.-T.}\
  \bibnamefont {Chen}}, \bibinfo {author} {\bibfnamefont {B.}~\bibnamefont
  {zhou}}, \bibinfo {author} {\bibfnamefont {G.-H.}\ \bibnamefont {Liang}},
  \bibinfo {author} {\bibfnamefont {Z.-C.}\ \bibnamefont {Xiang}}, \bibinfo
  {author} {\bibfnamefont {G.}~\bibnamefont {Xue}}, \bibinfo {author}
  {\bibfnamefont {D.}~\bibnamefont {Zheng}}, \bibinfo {author} {\bibfnamefont
  {K.}~\bibnamefont {Huang}}, \bibinfo {author} {\bibfnamefont {Z.-A.}\
  \bibnamefont {Wang}}, \bibinfo {author} {\bibfnamefont {H.}~\bibnamefont
  {Yu}}, \bibinfo {author} {\bibfnamefont {P.}~\bibnamefont {Sierant}},
  \bibinfo {author} {\bibfnamefont {K.}~\bibnamefont {Xu}},\ and\ \bibinfo
  {author} {\bibfnamefont {H.}~\bibnamefont {Fan}},\ }\bibfield  {title}
  {\bibinfo {title} {Many-body delocalization with a two-dimensional 70-qubit
  superconducting quantum simulator},\ }\bibfield  {journal} {\bibinfo
  {journal} {arXiv:2507.16882}\ }\href
  {https://doi.org/10.48550/arXiv.2507.16882} {10.48550/arXiv.2507.16882}
  (\bibinfo {year} {2025}{\natexlab{b}})\BibitemShut {NoStop}%
\bibitem [{\citenamefont {Kwon}\ and\ \citenamefont {Choi}()}]{Kwon2025}%
  \BibitemOpen
  \bibfield  {author} {\bibinfo {author} {\bibfnamefont {K.}~\bibnamefont
  {Kwon}}\ and\ \bibinfo {author} {\bibfnamefont {J.-y.}\ \bibnamefont
  {Choi}},\ }\bibfield  {title} {\bibinfo {title} {{\textit{In preparation}}},\
  }\href@noop {} {\ }\BibitemShut {NoStop}%
\bibitem [{\citenamefont {Sherson}\ \emph {et~al.}(2010)\citenamefont
  {Sherson}, \citenamefont {Weitenberg}, \citenamefont {Endres}, \citenamefont
  {Cheneau}, \citenamefont {Bloch},\ and\ \citenamefont {Kuhr}}]{Sherson2010}%
  \BibitemOpen
  \bibfield  {author} {\bibinfo {author} {\bibfnamefont {J.~F.}\ \bibnamefont
  {Sherson}}, \bibinfo {author} {\bibfnamefont {C.}~\bibnamefont {Weitenberg}},
  \bibinfo {author} {\bibfnamefont {M.}~\bibnamefont {Endres}}, \bibinfo
  {author} {\bibfnamefont {M.}~\bibnamefont {Cheneau}}, \bibinfo {author}
  {\bibfnamefont {I.}~\bibnamefont {Bloch}},\ and\ \bibinfo {author}
  {\bibfnamefont {S.}~\bibnamefont {Kuhr}},\ }\bibfield  {title} {\bibinfo
  {title} {{Single-atom-resolved fluorescence imaging of an atomic Mott
  insulator}},\ }\href {https://doi.org/10.1038/nature09378} {\bibfield
  {journal} {\bibinfo  {journal} {Nature}\ }\textbf {\bibinfo {volume} {467}},\
  \bibinfo {pages} {68–72} (\bibinfo {year} {2010})}\BibitemShut {NoStop}%
\bibitem [{\citenamefont {Anderson}(1958)}]{Anderson1958}%
  \BibitemOpen
  \bibfield  {author} {\bibinfo {author} {\bibfnamefont {P.~W.}\ \bibnamefont
  {Anderson}},\ }\bibfield  {title} {\bibinfo {title} {{Absence of Diffusion in
  Certain Random Lattices}},\ }\href {https://doi.org/10.1103/PhysRev.109.1492}
  {\bibfield  {journal} {\bibinfo  {journal} {Phys. Rev.}\ }\textbf {\bibinfo
  {volume} {109}},\ \bibinfo {pages} {1492–1505} (\bibinfo {year}
  {1958})}\BibitemShut {NoStop}%
\bibitem [{\citenamefont {Stöferle}\ \emph {et~al.}(2004)\citenamefont
  {Stöferle}, \citenamefont {Moritz}, \citenamefont {Schori}, \citenamefont
  {Köhl},\ and\ \citenamefont {Esslinger}}]{Stoferle2004}%
  \BibitemOpen
  \bibfield  {author} {\bibinfo {author} {\bibfnamefont {T.}~\bibnamefont
  {Stöferle}}, \bibinfo {author} {\bibfnamefont {H.}~\bibnamefont {Moritz}},
  \bibinfo {author} {\bibfnamefont {C.}~\bibnamefont {Schori}}, \bibinfo
  {author} {\bibfnamefont {M.}~\bibnamefont {Köhl}},\ and\ \bibinfo {author}
  {\bibfnamefont {T.}~\bibnamefont {Esslinger}},\ }\bibfield  {title} {\bibinfo
  {title} {{Transition from a Strongly Interacting 1D Superfluid to a Mott
  Insulator}},\ }\href {https://doi.org/10.1103/PhysRevLett.92.130403}
  {\bibfield  {journal} {\bibinfo  {journal} {Phys. Rev. Lett.}\ }\textbf
  {\bibinfo {volume} {92}},\ \bibinfo {pages} {130403} (\bibinfo {year}
  {2004})}\BibitemShut {NoStop}%
\bibitem [{\citenamefont {Kwon}\ \emph {et~al.}(2021)\citenamefont {Kwon},
  \citenamefont {Mukherjee}, \citenamefont {Huh}, \citenamefont {Kim},
  \citenamefont {Mistakidis}, \citenamefont {Maity}, \citenamefont
  {Kevrekidis}, \citenamefont {Majumder}, \citenamefont {Schmelcher},\ and\
  \citenamefont {Choi}}]{Kwon2021}%
  \BibitemOpen
  \bibfield  {author} {\bibinfo {author} {\bibfnamefont {K.}~\bibnamefont
  {Kwon}}, \bibinfo {author} {\bibfnamefont {K.}~\bibnamefont {Mukherjee}},
  \bibinfo {author} {\bibfnamefont {S.~J.}\ \bibnamefont {Huh}}, \bibinfo
  {author} {\bibfnamefont {K.}~\bibnamefont {Kim}}, \bibinfo {author}
  {\bibfnamefont {S.~I.}\ \bibnamefont {Mistakidis}}, \bibinfo {author}
  {\bibfnamefont {D.~K.}\ \bibnamefont {Maity}}, \bibinfo {author}
  {\bibfnamefont {P.~G.}\ \bibnamefont {Kevrekidis}}, \bibinfo {author}
  {\bibfnamefont {S.}~\bibnamefont {Majumder}}, \bibinfo {author}
  {\bibfnamefont {P.}~\bibnamefont {Schmelcher}},\ and\ \bibinfo {author}
  {\bibfnamefont {J.-y.}\ \bibnamefont {Choi}},\ }\bibfield  {title} {\bibinfo
  {title} {{Spontaneous Formation of Star-Shaped Surface Patterns in a Driven
  Bose-Einstein Condensate}},\ }\href
  {https://doi.org/10.1103/PhysRevLett.127.113001} {\bibfield  {journal}
  {\bibinfo  {journal} {Phys. Rev. Lett.}\ }\textbf {\bibinfo {volume} {127}},\
  \bibinfo {pages} {113001} (\bibinfo {year} {2021})}\BibitemShut {NoStop}%
\bibitem [{\citenamefont {Wienand}\ \emph {et~al.}(2024)\citenamefont
  {Wienand}, \citenamefont {Karch}, \citenamefont {Impertro}, \citenamefont
  {Schweizer}, \citenamefont {McCulloch}, \citenamefont {Vasseur},
  \citenamefont {Gopalakrishnan}, \citenamefont {Aidelsburger},\ and\
  \citenamefont {Bloch}}]{Wienand2024}%
  \BibitemOpen
  \bibfield  {author} {\bibinfo {author} {\bibfnamefont {J.~F.}\ \bibnamefont
  {Wienand}}, \bibinfo {author} {\bibfnamefont {S.}~\bibnamefont {Karch}},
  \bibinfo {author} {\bibfnamefont {A.}~\bibnamefont {Impertro}}, \bibinfo
  {author} {\bibfnamefont {C.}~\bibnamefont {Schweizer}}, \bibinfo {author}
  {\bibfnamefont {E.}~\bibnamefont {McCulloch}}, \bibinfo {author}
  {\bibfnamefont {R.}~\bibnamefont {Vasseur}}, \bibinfo {author} {\bibfnamefont
  {S.}~\bibnamefont {Gopalakrishnan}}, \bibinfo {author} {\bibfnamefont
  {M.}~\bibnamefont {Aidelsburger}},\ and\ \bibinfo {author} {\bibfnamefont
  {I.}~\bibnamefont {Bloch}},\ }\bibfield  {title} {\bibinfo {title}
  {{Emergence of fluctuating hydrodynamics in chaotic quantum systems}},\
  }\href {https://doi.org/10.1038/s41567-024-02611-z} {\bibfield  {journal}
  {\bibinfo  {journal} {Nat. Phys.}\ }\textbf {\bibinfo {volume} {20}},\
  \bibinfo {pages} {1732–1737} (\bibinfo {year} {2024})}\BibitemShut
  {NoStop}%
\bibitem [{\citenamefont {Hauschild}\ \emph {et~al.}(2024)\citenamefont
  {Hauschild}, \citenamefont {Unfried}, \citenamefont {Anand}, \citenamefont
  {Andrews}, \citenamefont {Bintz}, \citenamefont {Borla}, \citenamefont
  {Divic}, \citenamefont {Drescher}, \citenamefont {Geiger}, \citenamefont
  {Hefel}, \citenamefont {Hémery}, \citenamefont {Kadow}, \citenamefont
  {Kemp}, \citenamefont {Kirchner}, \citenamefont {Liu}, \citenamefont
  {Möller}, \citenamefont {Parker}, \citenamefont {Rader}, \citenamefont
  {Romen}, \citenamefont {Scalet}, \citenamefont {Schoonderwoerd},
  \citenamefont {Schulz}, \citenamefont {Soejima}, \citenamefont {Thoma},
  \citenamefont {Wu}, \citenamefont {Zechmann}, \citenamefont {Zweng},
  \citenamefont {Mong}, \citenamefont {Zaletel},\ and\ \citenamefont
  {Pollmann}}]{tenpy}%
  \BibitemOpen
  \bibfield  {author} {\bibinfo {author} {\bibfnamefont {J.}~\bibnamefont
  {Hauschild}}, \bibinfo {author} {\bibfnamefont {J.}~\bibnamefont {Unfried}},
  \bibinfo {author} {\bibfnamefont {S.}~\bibnamefont {Anand}}, \bibinfo
  {author} {\bibfnamefont {B.}~\bibnamefont {Andrews}}, \bibinfo {author}
  {\bibfnamefont {M.}~\bibnamefont {Bintz}}, \bibinfo {author} {\bibfnamefont
  {U.}~\bibnamefont {Borla}}, \bibinfo {author} {\bibfnamefont
  {S.}~\bibnamefont {Divic}}, \bibinfo {author} {\bibfnamefont
  {M.}~\bibnamefont {Drescher}}, \bibinfo {author} {\bibfnamefont
  {J.}~\bibnamefont {Geiger}}, \bibinfo {author} {\bibfnamefont
  {M.}~\bibnamefont {Hefel}}, \bibinfo {author} {\bibfnamefont
  {K.}~\bibnamefont {Hémery}}, \bibinfo {author} {\bibfnamefont
  {W.}~\bibnamefont {Kadow}}, \bibinfo {author} {\bibfnamefont
  {J.}~\bibnamefont {Kemp}}, \bibinfo {author} {\bibfnamefont {N.}~\bibnamefont
  {Kirchner}}, \bibinfo {author} {\bibfnamefont {V.~S.}\ \bibnamefont {Liu}},
  \bibinfo {author} {\bibfnamefont {G.}~\bibnamefont {Möller}}, \bibinfo
  {author} {\bibfnamefont {D.}~\bibnamefont {Parker}}, \bibinfo {author}
  {\bibfnamefont {M.}~\bibnamefont {Rader}}, \bibinfo {author} {\bibfnamefont
  {A.}~\bibnamefont {Romen}}, \bibinfo {author} {\bibfnamefont
  {S.}~\bibnamefont {Scalet}}, \bibinfo {author} {\bibfnamefont
  {L.}~\bibnamefont {Schoonderwoerd}}, \bibinfo {author} {\bibfnamefont
  {M.}~\bibnamefont {Schulz}}, \bibinfo {author} {\bibfnamefont
  {T.}~\bibnamefont {Soejima}}, \bibinfo {author} {\bibfnamefont
  {P.}~\bibnamefont {Thoma}}, \bibinfo {author} {\bibfnamefont
  {Y.}~\bibnamefont {Wu}}, \bibinfo {author} {\bibfnamefont {P.}~\bibnamefont
  {Zechmann}}, \bibinfo {author} {\bibfnamefont {L.}~\bibnamefont {Zweng}},
  \bibinfo {author} {\bibfnamefont {R.~S.~K.}\ \bibnamefont {Mong}}, \bibinfo
  {author} {\bibfnamefont {M.~P.}\ \bibnamefont {Zaletel}},\ and\ \bibinfo
  {author} {\bibfnamefont {F.}~\bibnamefont {Pollmann}},\ }\bibfield  {title}
  {\bibinfo {title} {{Tensor network Python (TeNPy) version 1}},\ }\href
  {https://doi.org/10.21468/SciPostPhysCodeb.41} {\bibfield  {journal}
  {\bibinfo  {journal} {SciPost Phys. Codebases}\ ,\ \bibinfo {pages} {41}}
  (\bibinfo {year} {2024})}\BibitemShut {NoStop}%
\end{thebibliography}%

\newcommand{\beginsupplement}{%
        \setcounter{table}{0}
        \renewcommand{\thetable}{S\arabic{table}}%
        \setcounter{figure}{0}
        \renewcommand{\thefigure}{S\arabic{figure}}%
        \setcounter{equation}{0}
        \renewcommand{\theequation}{S\arabic{equation}}%
     }
\clearpage
          
\newpage
\beginsupplement

\onecolumngrid 
\baselineskip18pt

\begin{center}
\large{\textbf{Supplemental Materials for \\ ``Stability of many-body localization in two dimensions"}}\\
\end{center}

\section{Experimental systems}\label{sec:1} 
\subsection{Experimental Sequence}\label{seq} 
As a first step to create the initial state, we prepared ultracold Bosonic $^{7}\text{Li}$ atoms in a 2D square optical lattice with a lattice constant of $a_{\rm lat} = 752$~nm~\cite{Kwon2022}. 
The vertical confinement is provided by an accordion lattice along $z-$axis, and the radial curvature is provided by an optical trap. 
Before ramping up the lattice depth to the Mott insulating regime, we apply a repulsive stripe-pattern potential, which is created by a digital micromirror device (DMD)~\cite{Fukuhara2013,Mazurenko2017}.
The stripe pattern has two-site periodicity ($\Lambda_{\text{Stripe}} = 2a_{\text{lat}}$), and a square wall boundary was imposed to confine the atoms in a finite system size $L\times{}L$~\cite{Kwon2025}.
Depending on the system size, we control the atom number by ramping down the radial trap, and then we transfer the atoms to the trap formed by the repulsive DMD potential. 
Finally, we ramp the lattice depth up to $V_{\text{lat}}=28.7 E_r$ to form a Mott insulator state in the presence of the patterning potential. 
The $E_r/h=12.6$~kHz is the recoil energy of the lattice, and $h$ is Planck's constant. 
Consequently, the atoms are arranged in an $n=1$ shell Mott insulator when the repulsive DMD potential is absent, whereas an $n=0$ state occurs when the DMD potential is present.
The Mott insulator state with stripe pattern at $L\times L = 24 \times 24$ was optimized to achieve unity filling $\bar{n} =0.91(4)$ by tuning the Feshbach resonance and DMD patterning beam power. 
This optimization process was performed for each system size to achieve the maximum initial state imbalance.
Furthermore, while taking data, we monitor the initial state every 5 minutes to obtain reliable measurements. We post-selected realizations with initial states characterized by sufficient imbalance $\mathcal{I}\geq 65-0.85 $, which depends on the system size.

Then the magnetic field was adjusted to $B=676.9\,\text{G}$, setting the atomic scattering length to $a=31a_B$ with $a_B$ being the Bohr radius. 
The dynamics is started by ramping down the two-dimensional optical lattice to $18E_r$, initiating atomic tunneling with a hopping energy of $J/h = 84(2)$~Hz. 
This configuration resulted in an on-site interaction energy of $U=478(6)$~Hz, yielding a $U/J =5.7(2)$.
Details about the parameter calibration are described in Section~\ref{cal}. 
The ramp-down time of the lattice was chosen to be 1~ms, which prevents unwanted inter-band heating during the quench process. 
During the evolution, the atoms interacted with a disorder potential generated by the DMD, where the potential realization was randomly chosen from a pre-generated disorder pattern realization. 
To ensure proper averaging over disorder realizations, we prepared 500 distinct disorder realizations, from which one was randomly selected for each experimental run. 
After the system evolved for various durations under these conditions, the lattice was rapidly ramped up again. 
Finally, the parity-projected occupation number of each site was measured using a quantum gas microscope~\cite{Kwon2022}.

\subsection{Programmable potential}\label{DMD}

\begin{figure}[ht]
\includegraphics[width=0.49 \linewidth]{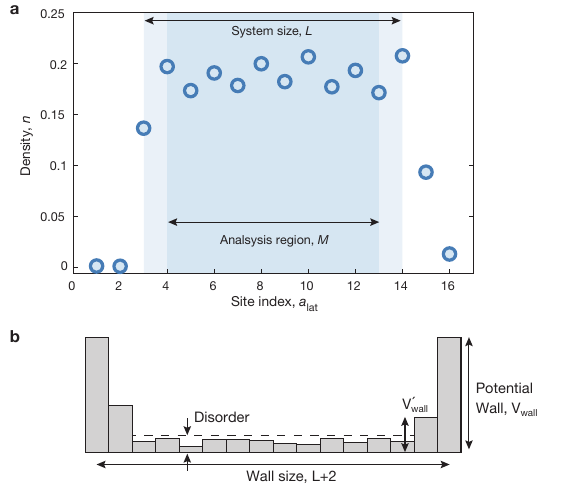}
\caption{
\textbf{Potential from digital micromirror devices.}
\textbf{a,} The averaged density of particles after an evolution time of $t_{\text{evol}} = 150\tau$ within a system of size $L \times L = 12 \times 12$. We applied a uniform random disorder with a potential of $W_\text{r}=64J$ to enter the MBL regime, where the initial stripe pattern is visible. 
\textbf{b,} A cross-section view of the potential landscape used in the experiments.
The finite optical resolution softens the boundary walls, having small potential offset. 
The analysis was restricted to a smaller interior region of $M \times M$ with $M=L-1$ to eliminate wall effects.
Random disorder potential is applied in the region $(L+1) \times (L+1)$.
} 
\label{mag-figS-WallInclusion}
\end{figure}

In our experiments, the key element is a programmable optical potential with single-site resolution, realized by projecting patterns from a digital micromirror device onto a two-dimensional square lattice through a high-magnification (150×) imaging system~\cite{Kwon2025}. 
Each lattice site corresponds to 8.25×8.25 DMD pixels, enabling grayscale potentials via binary dithering.
We use a blue-detuned 650 nm superluminescent diode as a light source, and amplified to have 35~mW after the fiber coupling. 
Residual phase drift of the optical lattice ($\sim$0.5 site/hour) is stabilized by monitoring lattice phase from fluorescence images and applying feedback to the DMD pattern at the end of each experimental run.
In this subsection, we provide a detailed description about the disorder for random and quasi-periodic potential, and the optical barrier.

\subsubsection{Disorder Potential} 
After preparing the initial state, we adjusted the system parameters for the main experiment while maintaining the lattice depth strong enough to prevent atomic tunneling. The magnetic field was tuned to modulate the on-site interaction potential so that the interaction-to-tunneling ratio is $U/J=5.7$. Simultaneously, the patterning potential was switched from the initial stripe pattern to a disorder pattern using an externally triggered transition. 
In the experiment, we use two different types of disorders.
First, uniform random disorder was generated from a quasirandom generator, with potential values $W_{{\bf i}=(x,y)}$ uniformly distributed within the range, $W_{\bf i}\in[0,W_\text{max}]$. 

Quasiperiodic Disorder was created by imposing another periodic potential with incommensurate period. 
The potential is given by
\begin{align}
W_{{\bf i}=(x,y)} =\frac{W_q}{4}\left[\cos(2\pi\beta_x x + \phi_x)+\cos(2\pi\beta_y y + \phi_y)\right].
\end{align}
Here, $\phi_x$ and $\phi_y$ are random phase offsets chosen independently for each experimental realization, and $\beta_x=0.721,\beta_y=0.693$. 
The strength of the disorder potential was controlled by either the intensity of the laser beam or the reflectivity of the DMD's mirrors. 
In addition to the disorder potential, we added the harmonic curvature compensating potential to the disorder pattern, which can deconfine the harmonic curvature caused by the optical lattice beam. 
The specific disorder configuration for each experiment was generated by external software and changed for each experimental run. 

\subsubsection{Potential barrier}

The overall pattern in the DMD is composed of a boundary wall and an interior disorder potential.  
The boundary walls height was set to maximum intensity, having the potential height of $V_{\rm wall}/h\geq15$~kHz, which is much larger than the tunneling energy $V_{\rm wall}/J\geq 180$. 
Because of the finite imaging resolution, however, we have one site broadening of the potential wall with its height $V'_{\rm wall}/h\geq6$~kHz.  
The maximum intensity of this interior disorder potential was intentionally set lower than that of the boundary walls, creating a potential gradient between the interior and exterior regions. 
Typically, the maximum intensity for the interior potential was $20-40\%$ of the DMD's maximum intensity. 
To avoid a homogeneous potential landscape at one site off from the wall, we also introduce the disorder pattern in this region. 
Overall, the disorder potential pattern is covered in the $(L+2)\times(L+2)$ region.

\subsubsection{Calibration of the disorder potential}\label{caldis}

\begin{figure}[b]
\includegraphics[width=0.49 \linewidth]{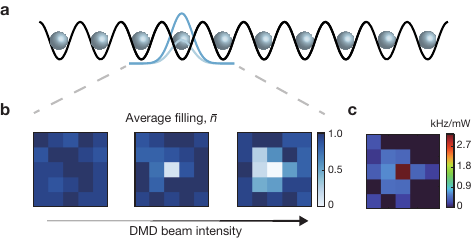}
\caption{
\textbf{Calibration of the DMD potential wall.}
\textbf{a,} A schematic illustration of the DMD potential calibration process. 
A single-site potential, generated by the DMD, is applied at target lattice sites.
The potential height is controlled by the light intensity for the DMD (blue curves).
\textbf{b,} The average atom density distributions during the superfluid to Mott insulator transition at different light intensities.
The DMD potential targets the central lattice site.
\textbf{c,} Calibrated potential height near the target site in terms of the total beam power applied to DMD. 
} 
\label{figS:DMD1}
\end{figure}

\begin{figure}[ht]
\centerline{\includegraphics[width=0.45\linewidth]{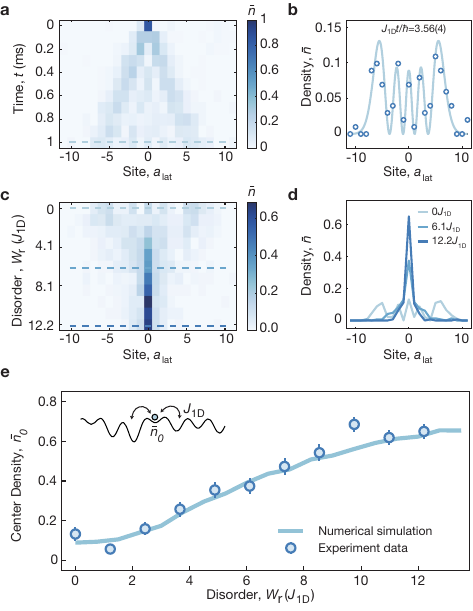}}
\caption{\textbf{Quantum walk under disorder potential.}
\textbf{a,} Density profile of a single particle as a function evolution time $t$ in a one-dimensional optical lattice.  
In the absence of disorder, the atomic wave front expands linearly over the time with interference pattern.
\textbf{b,} The interference density distribution  after at $t = 1~\text{ms}$ of quantum walk (blue empty symbols).
Solid line is a fit function to the experimental data, from which we can obtain the hopping constant, $J_{\rm 1D}$.
 \textbf{c,} Density profile after $t=1$ms of time evolution under various disorder strength. 
As the disorder strength increases (from top to bottom), an atom is localize near its initial position, displaying the Anderson localization.
\textbf{d,} Cross-section density profiles for various disorder strengths (dashed lines in c). 
In contrast to the interference pattern in the clean case ($W_{\rm r}=0J_{\rm 1D}$), increasing the disorder localizes the atom to its initial state, with the effect becoming more pronounced at higher disorder.
\textbf{e,} The density at the center ($\bar{n}_0$) is plotted as a function of disorder strength.
The experimental data (blue symbols) are in remarkable agreement with the exact diagonalization (ED) simulation (blue solid line).
} 
\label{mag-figS-DMDCalibQW}
\end{figure}

To characterize the disorder potential, we study the response of the atoms to the incident DMD beam power during the superfluid to Mott insulator transition and extract the point spread function of the optical projection system. 
We first prepared a Mott insulator without the DMD potential with an average density of $\bar{n} \approx 1$.
At unity filling Mott insulator in low temperature, the local density can approximate to $\bar{n}(x,y)=1/[1+e^{-\beta\mu(x,y)}]$, where $\beta=1/k_BT$ with $k_B$ is the Boltzmann constant, $\mu(x,y)=\mu_0-V_{\rm trap}(x,y)$ is the local on-site potential with the harmonic potential $V_{\rm trap}(x,y)$.
Fitting the radial profile of our Mott insulator, we determined the temperature $\beta$ and the local chemical potential $\mu(x,y)$~\cite{Sherson2010}. 
Then, we repeat the experiment with increasing DMD beam power, where the DMD light addresses a target lattice site (Fig.~\ref{figS:DMD1}a).
The density at a given site decreased with increasing DMD potential because the potential repels the atoms (Fig.~\ref{figS:DMD1}b).
The potential from the DMD light $V^{\rm DMD}(x,y)$ effectively reduces the local chemical potential to $\mu'(x,y)=\mu(x,y)-V_{\rm DMD}(x,y)$.
From the density profile at different beam intensity, we find the $V_{\rm DMD}(x,y)$ from the condition $\bar{n}=0.5$, which means $\mu'(x,y)=0$ or $V_{\rm DMD}(x,y)=\mu(x,y)$. 
Away from the target site, the density response on the addressing beam is hardly visible, and we set $V_{\rm DMD}(x,y)=0$ in such a case.

After determining the potential $V_{\rm DMD}(x,y)$ from the single-site addressing, we calibrated the single-site potential, as shown in Fig. \ref{figS:DMD1}c. 
By fitting the calibration data to a two-dimensional Gaussian function $V_{\rm DMD}(x,y)=V_0e^{-[(x-x_0)^2+(y-y_0)^2]/2\sigma^2}$, we obtained the broadening parameters from the finite optical resolution as $\sigma=0.7a_{\text{lat}}$, where $a_{\text{lat}}=752~\text{nm}$ is the lattice spacing. 
Also, the central potential in terms of the applied beam power was determined to be $V_0=2.8~\text{kHz}/\text{mW}$.
From this result, we can estimate disorder distribution in the atomic plane by convolving disorder images with $V_{\rm DMD}(x,y)$ (Fig.~\ref{mag-fig1} b,c). 
Here, the disorder strength is characterized by the full width at half maximum (FWHM) of the disorder distribution. 
With the reflectivity of the DMD being 0.4, the calibrations for the disorders are $W_{\rm r}=5.3~J/\text{mW}$ and $W_{\rm q}=3.3~J/\text{mW}$ for the uniform random disorder and the quasiperiodic disorder, respectively.

Furthermore, we perform a one-dimensional quantum walk under a random disorder potential and complete the disorder calibration. 
In this experiment, we prepared a single atom in a 1D optical lattice~\cite{Kwon2025}, where it shows a quantum random walk in a clean system (Fig. \ref{mag-figS-DMDCalibQW}a). 
We find the density distribution is well described by the equation, $n_l(t) = |\mathcal{J}_{l}(2J_{\rm{1D}}t/\hbar)|^2$, where $\mathcal{J}_l(x)$ is the Bessel function of the first kind, $l$ is the distance from the initial position, $J_{\text{1D}}$ is the 1D hopping energy, and $t$ is the evolution time. 
By fitting this function to the distribution at an evolution time of $t = 1~\text{ms}$, we extracted the hopping parameter of the 1D lattice system as $J_{\rm1D}/h = 570~\text{Hz}$ (Fig. \ref{mag-figS-DMDCalibQW}b).
Then, we apply a random uniform disorder to the 1D system, where Anderson localization is expected with increasing disorder strength~\cite{Anderson1958}. 
The localization is reflected in the density at the central position $n_0$, where the initial state information does not spread over the entire system and is localized  (Fig. \ref{mag-figS-DMDCalibQW} c, d). 
The experimentally measured average density $\bar{n}_0$ shows a good agreement with the numerical curve, in which the convoluted disorder distributions have been used.

\subsubsection{Correlation of disorders}

\begin{figure}[b]
\centerline{\includegraphics[width=0.45 \linewidth]{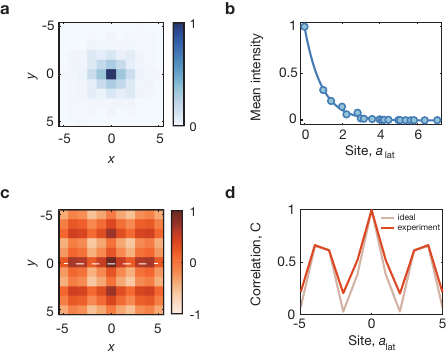}}
\caption{
\textbf{Two-point correlation function of disorder potentials.} 
\textbf{a,} The spatial correlation function for the uniform random disorder. 
\textbf{b,} Radially averaged profile of the correlation function. 
Solid line is an exponential fit curve with a decay length $\xi = 0.89(2)a_{\text{lat}}$. 
\textbf{c,} The correlation function for the quasiperiodic disorder. 
\textbf{d,} Cross-section view of the correlation over distance (dashed line in c). 
Because of the quasiperiodic pattern, it shows a long-range correlation. 
} 
\label{mag-figS-DisorderCorrelation}
\end{figure}

Characterizing our potential projection system, we are able to characterize the disorder potential used in the experiments. 
The random disorder potential, for example, is convoluted with the experimentally determined point spread function $V_{\rm DMD}$, and the homogeneous disorder distribution is changed to have a flat-top shape distribution (Fig.~1b).
Moreover, the finite resolution introduces a finite correlation length for the random disorder potential. 
The correlation length can be obtained by studying the two-point correlation function of the convoluted disorder potential $W_{\rm r,q}(x,y)$ with the lattice index $(x,y)$ in two dimensions, 
\begin{equation}
C(x,y)=\sum_{x',y'}\left[\langle W_{\rm r,q}(x,y)W_{\rm r,q}(x+x', y+y') \rangle -\langle W_{\rm r,q}(x,y)\rangle \langle W_{\rm r,q}(x+x', y+y')\rangle\right]
\end{equation}

For a uniform random disorder (Fig.~\ref{mag-figS-DisorderCorrelation}a), the spatial correlation function $C(x,y)$ was found to decay exponentially with distance. A fit to the exponential function $C(x,y) = C_0 \exp(-\sqrt{x^2+y^2}/\xi)$ (Fig.~\ref{mag-figS-DisorderCorrelation}b) yielded a correlation length of $\xi = 0.89(2)a_{\text{lat}}$. 
In contrast, the spatial correlation for the quasiperiodic disorder (Fig.~\ref{mag-figS-DisorderCorrelation}c,d) doesn't follow a simple exponential decay. Instead, it shows a clear long-range periodic correlation from its quasiperiodic nature. The finite optical resolution of the system also affects this correlation, smoothing out the distribution as seen in Fig.~\ref{mag-figS-DisorderCorrelation}d.

\subsection{Hubbard parameters}\label{cal}

\begin{figure}[b]
\centerline{\includegraphics[width= \linewidth]{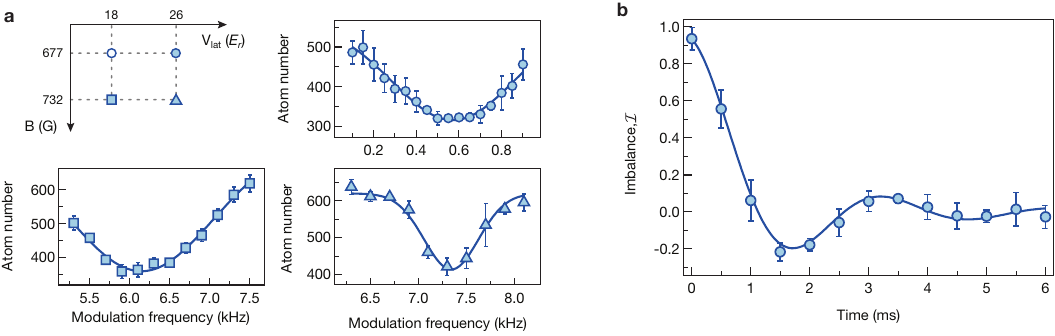}}
\caption{\textbf{Calibrating interaction and tunneling energy.}
\textbf{a,} Calibrating onsite interaction energy $U$. 
The experiment is carried out at the lattice depth $V_{\text{lat}} = 18E_r$ and a magnetic field of $B=676.9\,\text{G}$ (open circle), where the tunneling energy is non-negligible. 
We tune our parameters in the regime with large $U/J\gg1$, which is achieved by either increasing lattice depth or scattering length (three different symbols). 
In each regime, we perform the modulation spectroscopy, extracting the onsite interaction energy $U$.
Solid lines represent a Gaussian fit to the data.
\textbf{b,} Time evolution of the imbalance for strongly-interacting bosons in two dimensions, initiated with a stripe density pattern.
The tunneling energy is obtained by fitting the imbalance with $\mathcal{I}(t) = \mathcal{I}_0e^{-\gamma t}\mathcal{J}_0(4Jt)$, where $I_0$ is the initial imbalance, $\mathcal{J}_0$ is the Bessel function of the first kind, $\gamma$ is a decay constant, and $J$ is the tunneling parameter.
} 
\label{mag-figS-JU}
\end{figure}

The on-site interaction energy $U$ was measured using modulation spectroscopy~\cite{Stoferle2004}. 
While this technique is robust, its validity is contingent upon the tunneling parameter $J$ being sufficiently small compared to the on-site interaction $U$. 
Unfortunately, our experimental parameter $U/J=5.5$, did not meet this prerequisite for a direct measurement via modulation spectroscopy. 
Therefore, to accurately determine the on-site interaction energy $U$ for our experimental parameters, we first measure the interaction energy in the large $U/J$ limit at large scattering length or deep lattice and then estimate the interaction energy at the experimental condition.
We first measured $U$ at a large lattice depth of $V_{\text{lat}} = 26E_r$, where the tunneling is $J/h\simeq 23$~Hz.
From the modulation spectroscopy, we find the onsite interaction energy $U/h=7.34(2)$~kHz and $U/h=571(7)$~Hz at two different magnetic fields $B_1=732$~G and $B_2=676.9$~G, respectively. 
Subsequently, using the same method, we measured the on-site energy at a shallower lattice depth $V_{\text{lat}} = 18E_r$ to be  $U/h=6.15(2)$~kHz. From these results, we can infer the interaction energy at $B_2$ at $18E_r$ as $U/h=478(6)$~Hz.

The hopping parameter $J$ is obtained from the dynamics of hard-core bosons in two dimensions, where the strong on-site interaction is obtained by ramping up the Feshbach magnetic field to have $a_{\text{lat}} =1100a_B$~\cite{Kwon2021}. 
After preparing the stripe pattern as described in Section~\ref{seq}, we ramp down the lattice depth to $V_{lat}=18E_r$, and study the time evolution of the atoms without a disorder potential. 
Fig.~\ref {mag-figS-JU}b displays the dynamics of the imbalance, and we observe a Bessel-like oscillation even with exponential decay, which is consistent with the recent experiment in a 2D homogeneous optical lattice system~\cite{Wienand2024}. 
From the fitting of the oscillation, we obtained the tunneling energy to be $J/h = 84(2)\,\text{Hz}$.

\subsection{Temperature after time evolution}
\begin{figure}[ht]
\centerline{\includegraphics[width=0.45 \linewidth]{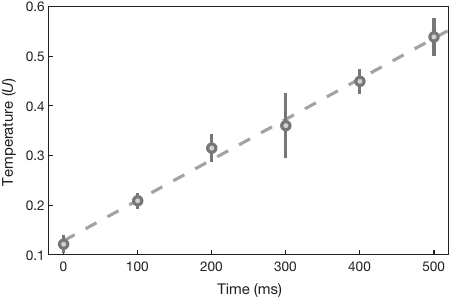}}
\caption{\textbf{Heating effect.} Temperature as a function of hold time in the optical lattice at the lattice  $V_{\text{lat}} = 15E_r$. 
The dashed line is a linear fit with the heating rate, ${dT}/{dt}=0.8U/s$.} 
\label{mag-figS-Temperature}
\end{figure}

During the experimental dynamics, atoms can acquire heat from various external sources, including pointing and phase fluctuations of the optical lattice, as well as collisions with residual vacuum atoms and blackbody radiation. 
To assess this heating effect, we measured the system's temperature near the critical point ($V_{\text{lat}} = 15E_R$), where the system is most susceptible to external thermal perturbations. 
In this measurement, the DMD potential is not applied, and the temperature is obtained from the radial density distribution as discussed in Sec.~\ref{caldis}.
As shown in Fig.~\ref{mag-figS-Temperature}, a temperature increases at a rate of ${dT}/{dt} = 0.8U/s$.

\section{Numerical simulations}\label{sec:2} 
Complementing the experimental work, we also performed numerical simulations using the time-dependent variational principle (TDVP) on matrix product states (MPS)~\cite{tdvp}. We followed the methods of Ref.~\cite{Doggen2020} and used the TeNPy library~\cite{tenpy}. We considered $L \times L$ square lattices with open boundary conditions, focusing on the system sizes $L=6, 8$ and 20 disorder realizations in each case. Following a similar protocol as in the experiment, we initialized the system in a charge-density-wave state with alternating full and empty columns and simulated real time dynamics to calculate the imbalance as a function of time. Transition points were extracted by performing a power law fit $\mathcal{I}(t) = A\, t^{-b}$ to the long-time imbalance $\mathcal{I}(t) = (n_\mathrm{odd} - n_\mathrm{even})/(n_\mathrm{odd} + n_\mathrm{even})$. We note that this fit is only valid in the thermal regime (resulting in a positive $b$) as opposed to the MBL regime, where the imbalance tends to oscillate around high values (giving rise to a negative $b$). 

\begin{figure}[t]
\centerline{\includegraphics[width=\linewidth]{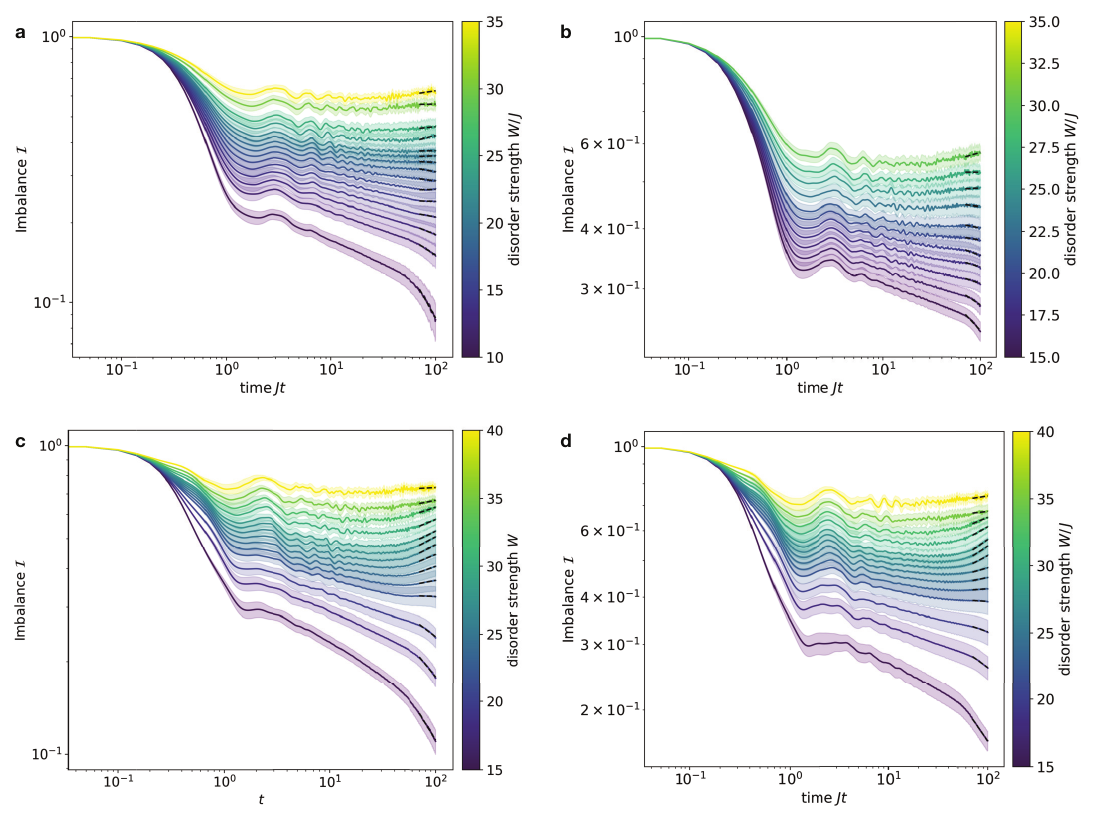}}
\caption{
Numerical simulations of the imbalance $\mathcal{I}(t)$ for the Hamiltonian in Eq.~\eqref{eq:ham}  for \textbf{a,} a $6 \times 6$ system with uniform random disorder, \textbf{b,} an $8 \times 8$ system with uniform random disorder, \textbf{c,} a $6 \times 6$ system with quasiperiodic random disorder, and \textbf{d,} an $8 \times 8$ system with quasiperiodic  disorder. 
} 
\label{simulationImbalance}
\end{figure}

\begin{figure}[t]
\centerline{\includegraphics[width= \linewidth]{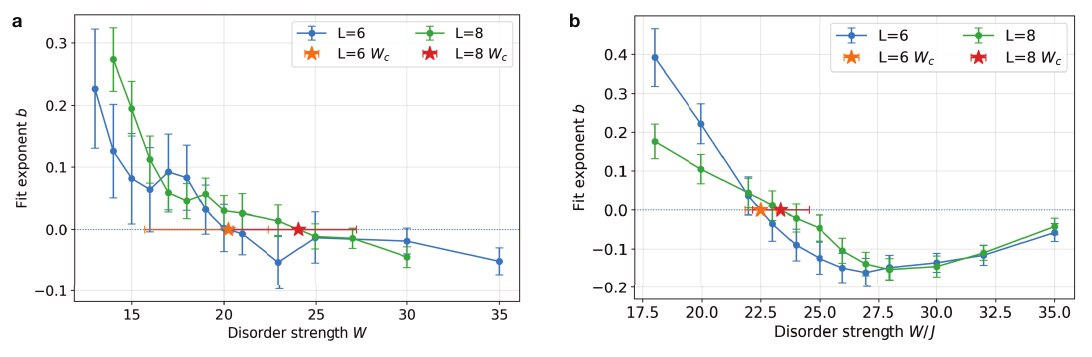}}
\caption{
Determination of the critical disorder strength for \textbf{a,}~uniform random disorder and \textbf{b,}~quasiperiodic disorder from the decay exponent $b$, obtained by fitting $\mathcal{I}(t) = A\, t^{-b}$ to the numerical simulation data in Fig.~\ref{simulationImbalance} for $L = 6$ and $L = 8$.
}
\label{simulation_b}
\end{figure}

The time evolution was generated by the Hamiltonian in Eq.~\eqref{eq:ham} for $J = 1$, with the local Hilbert space restricted to the three lowest on-site occupation numbers (0, 1, and 2). As in the experiment, we considered both random and quasiperiodic disorder, with the disorder potential $W_{\bf i}$ as defined below Eq.~\eqref{eq:ham}.
At each time step, the imbalance 
 was evaluated without applying the parity projection.
We found that for an MPS bond dimension of $\chi = 340$ and time step size $\Delta t = 0.05$, the results for the imbalance for $L = 8$ were sufficiently well converged, even for long evolution times of $t = 100 J^{-1}$ hoppings. 
Figure~\ref{simulationImbalance} shows the time dependence of the imbalance $\mathcal{I}(t)$ for random and quasiperiodic disorder. The data are fit to $\mathcal{I}(t) = A\, t^{-b}$ for evolution times $t \geq 70$. Figure~\ref{simulation_b} plots the extracted decay exponent $b$ as a function of disorder strength for the two disorder types. The transition points are obtained from the disorder strengths where $b$ becomes zero within error bars. We found that for the accessible system sizes of $L=6$ and $L=8$, the obtained critical disorder strengths of $W_{c,\mathrm{r}}(L = 6) = 20.2 \pm 4.5 J$, $W_{c,\mathrm{r}}(L = 8) = 24.1 \pm 3.6 J$, $W_{c,\mathrm{q}}(L = 6) = 22.5 \pm 0.7 J$, and $W_{c,\mathrm{q}}(L = 8) = 23.4 \pm 1.2 J$ are consistent with the experimental data. 

In the quasiperiodic case, we observe that $b$ is significantly smaller than zero for $W > W_{c,\mathrm{q}}(L)$. As discussed above, $\mathcal{I}(t) = A t^{-b}$ is not a reliable fit in the MBL regime, and we therefore identify the transition point as the value of $W$ where $b$ first reaches zero with increasing $W$. The negative values of $b$ also indicate that the feasible simulation times remain insufficient to reach the true long-time regime. To ensure numerical reliability, we also performed convergence checks and confirmed that the imbalance dynamics and associated short-term oscillations are stable with respect to increasing bond dimension. Although full access to the asymptotic regime is computationally unfeasible, the observed persistence of a finite imbalance at late times is consistent with the phenomenology of many-body localization.

\onecolumngrid 
\baselineskip18pt

\end{document}